\documentclass[runningheads]{llncs}

\usepackage{eccv}

\usepackage{eccvabbrv}

\usepackage{graphicx}
\usepackage{booktabs}

\usepackage{multirow}
\usepackage{tabularx} 
\usepackage{enumitem}

\usepackage{booktabs}  % 三线表
\usepackage{xcolor}    % 颜色支持
\usepackage{array}     % 表格增强

\usepackage{marvosym}

\usepackage[accsupp]{axessibility}  % Improves PDF readability for those with disabilities.

\usepackage{hyperref}

\usepackage{orcidlink}

\begin{document}

\title{Coding with Eyes: Visual Feedback Unlocks Reliable GUI Code Generating and Debugging} 

\titlerunning{Visual Feedback Unlocks Reliable GUI Code Generating and Debugging}

\author{
Zhilin Liu\inst{1} \and
Ye Huang\inst{1}%\textsuperscript{\Letter}
\orcidlink{0000-0001-5668-5529} \and
Ting Xie\inst{1} \and\\
Ruizhi Zhang\inst{1} \and
Wen Li\inst{1} \and
Lixin Duan\inst{1}\orcidlink{0000-0002-0723-4016}
}

\authorrunning{Zhilin Liu et al.}

\institute{Shenzhen Institute for Advanced Study,\\ University of Electronic Science and Technology of China
%\email{lncs@springer.com}
}

\setlength{\textfloatsep}{4pt}
\setlength{\floatsep}{4pt}
\setlength{\intextsep}{1pt}
\setlength{\abovecaptionskip}{1pt}
\setlength{\belowcaptionskip}{1pt}

\maketitle
\begin{abstract}
Recent advances in Large Language Model (LLM)-based agents have shown remarkable progress in code generation.
However, current agent methods mainly rely on text-output-based feedback (\eg command-line outputs) for multi-round debugging and struggle in graphical user interface (GUI) that involve visual information.
This is mainly due to two limitations: 1) GUI programs are event-driven, yet existing methods cannot simulate user interactions to trigger GUI element logic.
2) GUI programs possess visual attributes, making it difficult for text-based approaches to assess whether the rendered interface meets user needs.
To systematically address these challenges, we first introduce InteractGUI Bench, a novel benchmark comprising 984 commonly used real-world desktop GUI application tasks designed for fine-grained evaluation of both interaction logic and visual structure. 
Furthermore, we propose VF-Coder, a vision-feedback-based multi-agent system for debugging GUI code.
By perceiving visual information and directly interacting with program interfaces,  VF-Coder can identify potential logic and layout issues in a human-like manner.
On InteractGUI Bench, our VF-Coder approach increases the success rate of Gemini-3-Flash from 21.68\% to 28.29\% and raises the visual score from 0.4284 to 0.5584, indicating the effectiveness of visual feedback in GUI debugging. 
\end{abstract}    
\section{Introduction}
\label{sec:intro}

In recent years, the rapid progress of LLM-based agents has substantially advanced automated code generation. 
Models now perform competitively on widely used benchmarks such as HumanEval and MBPP\cite{chen2021humaneval,austin2021mbpp}, and have begun to power practical development assistants like Copilot and Cursor. 
However, effectively generating and debugging GUI code remains a challenging problem.

 \begin{figure}[t]
  \centering
  \includegraphics[width=0.9\linewidth, trim=9.5cm 5cm 9.2cm 2.5cm, clip]{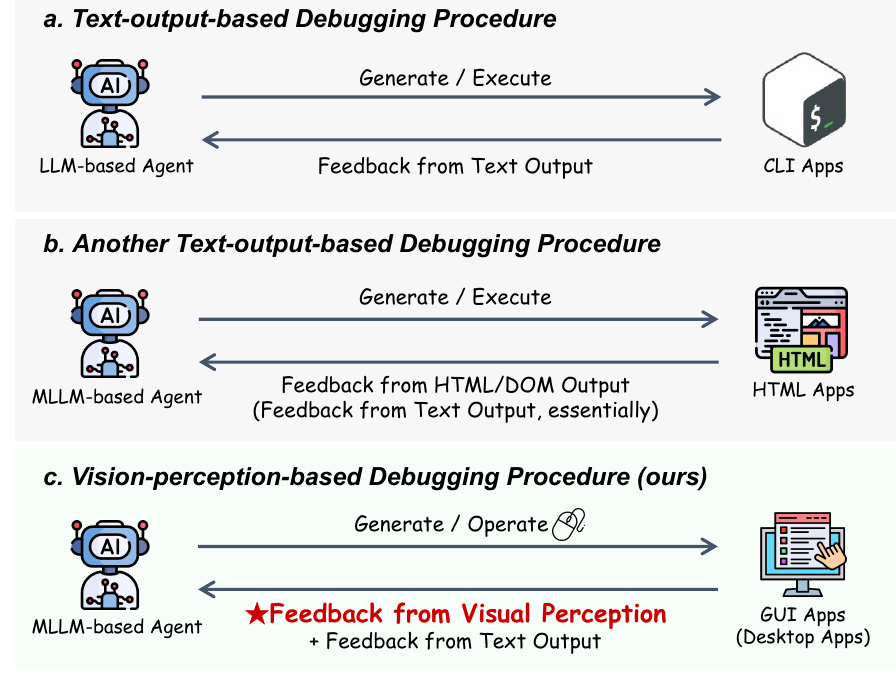}
  \caption{Comparison between (a, b) traditional text-output-based debugging procedure and (c) our vision-perception-based debugging procedure (VF-Coder), which enables direct perception of rendered GUIs for interactive correction.}
  \label{fig:singlecol}
\end{figure}

Existing code agents (\ie Fig.~\ref{fig:singlecol}.a) primarily rely on textual feedback for self-repair, iteratively optimizing code through the textual information like execution logs or error messages.
Some studies further incorporate external knowledge sources\cite{wang2025toolgen,Zhang2024CodeAgent}or 
structured information 
(\eg abstract syntax trees or XML Trees)\cite{Jiang2025RECODE,gupta2024codenav,zhang2025cast} for better visual effects. 
However, the absence of visual perception regarding the actual rendered interface prevents the code agent from accurately assessing layout and rendering correctness.

Although the advancement of Multimodal Large Language Models (MLLMs) offers visual perception for code agents,
making them now able to generate GUI code directly from real UI screenshots or design drafts for front-end development tasks (\ie Fig.~\ref{fig:singlecol}.b),
the debugging stage of these methods still relies on textual feedback such as UI layout trees\cite{wu2025layoutcoder} or DOM structures\cite{Yuan2025DesignRepair}.

This implies that the corrective capabilities of code agents remain confined to the textual level, lacking the ability to directly harness visual information for end-to-end dynamic debugging.
Furthermore, existing benchmarks and methods focus mainly on the visual reproduction of static webpages, and difficult to migrate to other GUI forms, such as Desktop GUI apps.
Also, existing benchmarks struggle to handle complex multi-page structures and long-range interaction logic in real-world scenarios.
%
%Consequently, they struggle to handle complex multi-page structures and long-range interaction logic in real-world scenarios and are difficult to migrate to other GUI forms, such as Desktop GUIs.
%
Therefore, these limitations reveal two core challenges in generating real-world GUI code: 
\begin{enumerate}[leftmargin=*, label=(\arabic*)]
\item \textbf{Functional logic modeling:}  
GUI apps follow an event-driven paradigm that emphasizes interactivity. 
Only user operations (\eg clicks, text inputs, or focus changes) can trigger behavioral or state changes, and varying interaction sequences may lead to diverse interface states and logical outcomes. 
This makes it difficult for models to capture dynamic dependencies and maintain functional correctness across interactions.

\item \textbf{Visual structural alignment:} 
Unlike command-line interface (CLI) programs that lack a visual component, the usability of a GUI is tightly coupled with its visual organization. 
The generated interface must not only be functionally correct but also preserve consistent layout structure, component styling, and visual hierarchy.
\end{enumerate}

To address these challenges, we first introduce \textbf{InteractGUI Bench}, a high-quality benchmark tailored for real-world desktop GUI scenarios.
We observed that existing datasets and benchmarks focus predominantly on Web scenarios, ignoring the diverse layout challenges and limited debugging information inherent in desktop GUI apps.
Thus, we carefully curated up to 984 diverse real-world desktop GUI apps as test cases, covering nearly all commonly used apps available on the Internet.
Each case consists of multiple interface screenshots and a textual instruction, reflecting the authentic workflow of human developers building GUIs from design drafts.

Furthermore, unlike most existing benchmarks that solely evaluate static interfaces, InteractGUI Bench establishes a fully executable sandbox environment and introduces Interactive Evaluation Script (IES), which can simulate real interactions and interface transitions on actually running GUI apps. 
Backed by the Assistive Technology Service Provider Interface (AT-SPI) and a specially trained visual evaluation model, we provide a comprehensive assessment ranging from atomic attributes (components, colors, interactions), precise layout, all the way to overall GUI structure.
We systematically evaluated mainstream MLLMs on InteractGUI Bench.
Taking PySide6 as an example, even state-of-the-art models achieve only a 26.99\% task success rate, showing the difficulty and complexity of our benchmark.

In addition, we propose \textbf{VF-Coder}, a multi-agent framework that integrates visual feedback into the entire lifecycle of GUI code generation and debugging.
Inspired by GUI Agent methods~\cite{zhou2025maiui, wang2025uitars2,nguyen2024guiagentsurvey}, VF-Coder allows the model to directly observe and manipulate rendered interfaces to detect and repair potential logical or display errors in real time.
On InteractGUI Bench, using Gemini-3-Flash as the base model, VF-Coder achieves a 6.61\% improvement in task success rate (from 21.68\% to 28.29\%), significantly outperforms existing text-based methods, which achieve only about a 3\% improvement. 
This result demonstrates the effectiveness of visual feedback in enhancing the quality of GUI apps. 

Our main contributions are as follows:
\begin{enumerate}
    \item We systematically identify the two fundamental challenges in GUI code generation: functional logic modeling and visual structural alignment. 
    Both capture the dual difficulties of maintaining interactivity and visual consistency.
    \item We introduce InteractGUI Bench, a high-quality benchmark comprising 984 real-world desktop GUI apps equipped with an end-to-end automated evaluation framework. 
    Leveraging AT-SPI and the visual evaluation model, it enables comprehensive measurement and accurate judgment.
    \item We propose VF-Coder, the first visual feedback-driven framework that introduces GUI Agent methodologies to solve the code generation task. 
    By simulating the real debugging process of human developers, VF-Coder establishes a closed-loop mechanism of ``visual perception - dynamic interaction - code refactoring'', achieving the automated repair of functional logic errors and visual rendering flaws, significantly enhancing the robustness and visual consistency of generated apps. 
\end{enumerate}
\section{Related Work }
\label{sec:related_work}

\subsection{Benchmarks for Code Generation }

Early benchmarks mainly evaluated models’ capabilities at the line or function level of code generation\cite{chen2021humaneval, austin2021mbpp, hendrycks2021apps, li2022CodeContests, jain2025livecodebench}.
With the rapid advancement of LLMs, many research works have expanded toward real-world app development tasks.

For example, SWE-Bench\cite{jimenez2024swebench} and CodeAgentBench\cite{Zhang2024CodeAgent} require models to perform functionality repair and feature development directly within large codebases.
EvoCodeBench\cite{JLHL2024evocodebench} and DevEval\cite{Li2024deveval} further evaluate whether models can generate target code that meets specified requirements under full project context and constraints.
Web-Bench\cite{xu2025webbench} extends to web development scenarios, emphasizing the agent’s capacity for long-term context understanding across multi-step development processes.
However, these benchmarks remain constrained to text-output-based feedback, thereby ignoring the inherent interactivity and visual characteristics of GUI environments.

Another line of research attempts to evaluate visual aspects using the image-to-code paradigm.
For instance, WebSight\cite{laurenccon2024websight} synthesizes two million HTML screenshot pairs, providing a large-scale foundation for studying visual code generation.
However, its code similarity metrics may cause models to overfit superficial patterns rather than true visual semantics.

To better capture front-end design fidelity, Design2Code\cite{SiZLYLY2025Design2Code} measures performance via visual similarity between generated pages and reference images.
Nonetheless, such methods focus solely on static appearance while ignoring end-to-end interaction.

To move beyond static evaluation, Interaction2Code\cite{xiao2024interaction2code} uses webpage screenshots before and after user interactions to simulate state transitions, encouraging models to reason about component functionality.
Yet, its dual-image differencing mechanism can only represent simple single-step interactions, falling short of capturing the multi-stage state dependencies and complex logic of real GUI applications. 
Overall, existing image-to-code benchmarks primarily focus on visual consistency in web front-end generation, but they lack systematic evaluation of multi-stage interaction logic and desktop-level GUI programming.

\subsection{Code Agents }

Recent LLM-based code generation agents primarily rely on text-output-based feedback for iterative refinement~\cite{self-edit,ye2025adverintent_agent,sohrabizadeh2025nemotroncortexa,antoniades2025swesearch}, leveraging execution outputs~\cite{self-edit} and other external knowledges~\cite{Manish2025autopatch} to perform self-correction and substantially improve generation accuracy.
For example, CodeNav~\cite{gupta2024codenav} automatically retrieves relevant code snippets from real repositories and debugs them based on execution results.
ROCODE~\cite{Jiang2025RECODE} continuously monitors compilation logs during generation and backtracks once errors are detected.
QualityFlow~\cite{hu2025qualityflow} adopts a multi-agent workflow that generates unit tests to more comprehensively verify code functionality.
However, these methods are not tailored for GUI-oriented scenarios and thus fail to model the visual and interactive characteristics inherent to GUI.

Several studies have begun to explore the use of visual information.
However, they typically incorporate it only once at the input stage, while the debugging process remains dominated by textual feedback.
For instance, LayoutCoder\cite{wu2025layoutcoder} constructs structured UI layout trees to enhance LLMs’ understanding of interface hierarchy and spatial arrangement, whereas DesignRepair\cite{Yuan2025DesignRepair} leverages rendered DOM trees to identify and repair visual defects.

Although these methods improve the visual consistency between generated code and reference screenshots, the lack of runtime visual perception and interaction prevents them from dynamically reasoning about rendered interfaces.
As a result, they are limited to generating static UI and struggle to handle GUI programs involving multi-page navigation and complex event-driven logic.

Our work distinguishes itself by incorporating visual feedback–driven vulnerability awareness and interactive debugging. 
By performing real interactive testing through screenshots, our method enables the agent to efficiently diagnose and fix program errors.

\section{Proposed InteractGUI Bench}
\label{sec:bench}

 \begin{figure*}[t]
  \centering
  \includegraphics[width=\linewidth, trim=0.2cm 1.6cm 0.3cm 0cm, clip]{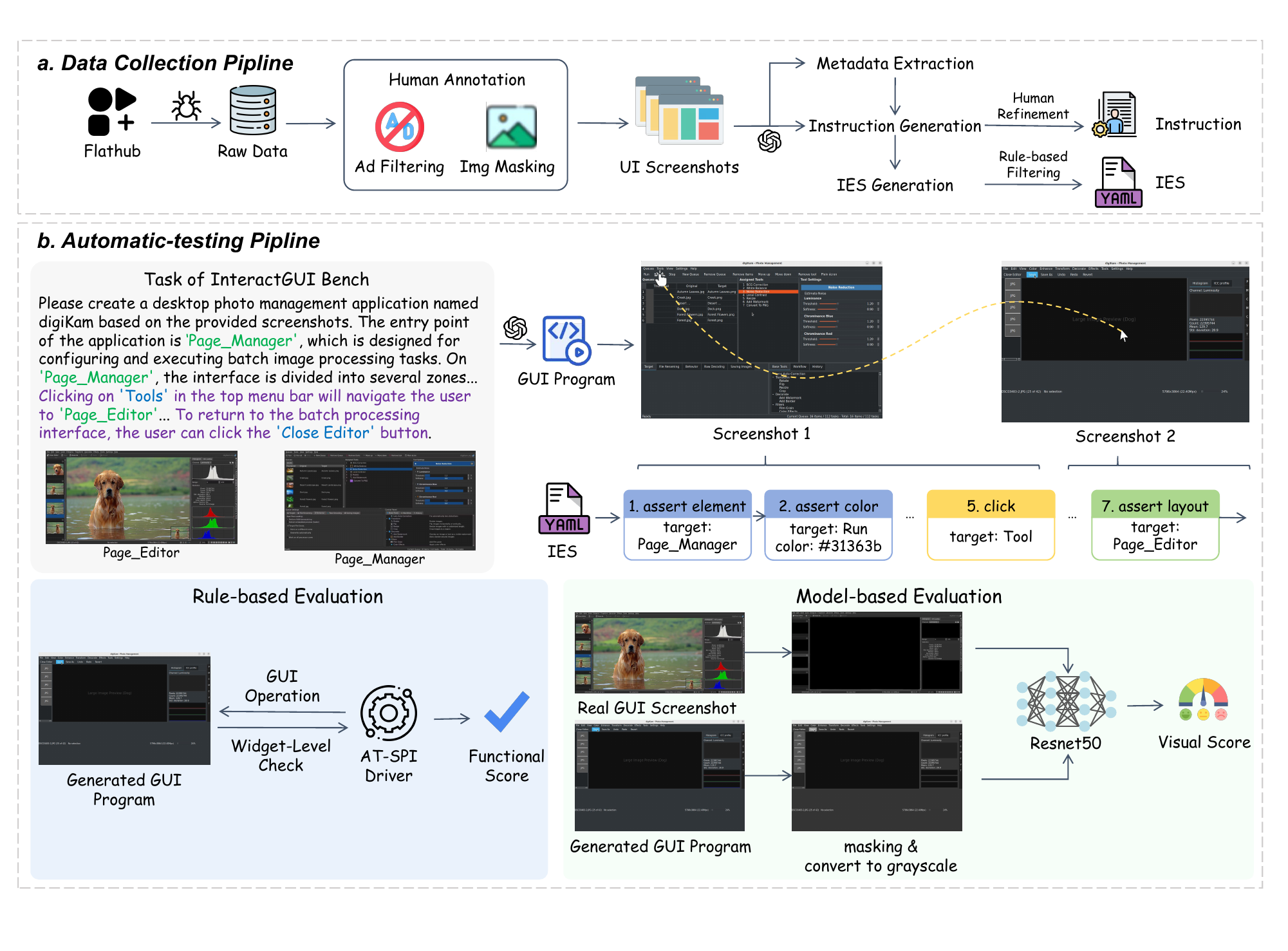}
  \caption{The data collection and automatic-testing pipeline. (a) shows the process of collecting real-world desktop UI screenshots from Flathub, constructing task instructions, and building the IES. (b) illustrates the workflow for testing generated GUI apps using a series of steps defined by the IES. Zoom in to see better.}
  \label{fig:data_collection_pipeline}
\end{figure*}

This section introduces the proposed InteractGUI Bench, a high-quality benchmark that consists of 984 GUI code generation tasks from 984 real-world desktop GUI apps.
Each task comprises multiple GUI screenshots of a real-world Desktop application and corresponding natural-language instructions. 
We first describe the data acquisition and task construction pipeline, followed by a detailed explanation of the automated verification and evaluation process. 

\subsection{Data Collection}

\subsubsection{Raw Data Acquisition:}
To ensure the authenticity and diversity of the benchmark, we collected data from Flathub, a centralized app store for Linux.
Flathub encompasses thousands of desktop applications, each with high-resolution, real-world GUI screenshots.
Using a Python-based crawler, we retrieved the top 1200 GUI apps from the ``Popular'' list, capturing their descriptions and corresponding interface screenshots as the foundational data source. 

\subsubsection{Manual Cleaning and Annotation:}
Due to the broad range of sources, the raw data contained significant noise.
We addressed two key challenges through manual review:
\begin{itemize}
    \item Filtering Non-GUI Content: Some application screenshots included promotional posters or commercial advertisements, which could cause model inference to deviate from the target. These were manually excluded to ensure data purity.
    \item Resource Sensitivity Handling: MLLMs cannot autonomously generate logos or images. To prevent such uncontrollable factors from biasing the visual evaluation, we manually performed masking on all icons and images within the screenshots, allowing the assessment to focus on visual structure and interaction logic.
\end{itemize}
Following this refinement, we distilled the initial 1200 candidate apps into 984 high-quality instances.
Each instance contains one or more GUI screenshots of the same desktop app.

\vspace{-3mm}

\subsubsection{Metadata Extraction and Instruction Generation:}
To transform static screenshots into executable tasks, we developed a multi-stage synthesis pipeline using Gemini-3-Flash~\cite{deepmind2025gemini_3_flash}.
We first prompt the model to analyze collected screenshots and descriptions, extracting structured metadata that includes component name and navigation logic between each interfaces. 
Based on this metadata, we then use model to generate natural language instructions that mimic human intent.
The instructions emphasize cross-page interaction, effectively bridging multiple independent screenshots into a unified GUI development task . 

\subsection{Evaluation}

\subsubsection{Construction of Interactive Evaluation Script:}

To support the evaluation of multi-interface layouts and interaction logic of generated GUI apps, we design the Interactive Evaluation Script (IES) for each task.
IES is a YAML-formatted declarative interaction protocol that describes interface states, user operations, and expected outcomes in a structured form.
We define the following three types of assertion operations and three types of interaction operations.
By combining these operations in IES, we enable dynamic verification and interaction for the generated GUI apps.

The assertion operations are used to verify visual and semantic consistency:
\begin{itemize}
    \item \textbf{Assert Element:} Specifies the expected components that should appear on the current page, which is used to verify the structural completeness of the interface and detect missing or misplaced elements.
    \item \textbf{Assert Color:} Specifies the expected color and style attributes of key visual elements, used to assess visual consistency and conformity with design specifications.
    \item     \textbf{Assert Layout:} Evaluates the visual structural similarity between a specified page and its corresponding reference UI screenshot.
\end{itemize}

The interaction operations, on the other hand, simulate concrete user interactions:
\begin{itemize}

    \item \textbf{Click:} Declares a click action on a specific component to simulate clicking behavior and evaluate its interactivity.
    \item \textbf{Input Text:} Declares a text-input operation within the target input field.
    \item \textbf{Dropdown Selection:} Declares a user selection operation from a dropdown menu.

\end{itemize}

Regarding the construction pipeline, we first utilize Gemini-3-Flash to generate a draft IES based on the task instruction.
Subsequently, a rule-based script performs consistency validation: each step defined in the IES is precisely matched against the Metadata and instruction, while strictly constraining interaction operations to occur only on components with navigation attributes.
This rigorous process ensures the reliability of the IES, enabling closed-loop verification of both visual rendering and interaction logic for the generated GUI apps.

\subsubsection{Rule-based Evaluation:}
We leverage the Assistive Technology Service Provider Interface (AT-SPI) to achieve component-level access and manipulation. 
AT-SPI is a Linux-supported system interface that provides access to the internal structure of GUI apps through the accessibility tree. 
This allows us to not only retrieve metadata such as component positions, hierarchies, states, and visibility to support the  \textbf{Assert Element} operation but also to execute \textbf{Click}, \textbf{Input Text} and \textbf{Dropdown Selection} operations.

For \textbf{Assert Color}, since AT-SPI does not directly expose color or style attributes, we further utilize the component coordinates it provides to capture the corresponding screen regions at runtime.
We then perform a pixel-wise color comparison by calculating the Euclidean distance $d$ in the RGB color space between the rendered pixels and the target color; a threshold of $d < 80$ is applied to determine if the assertion passes. 

\subsubsection{Model-based Evaluation:}
For \textbf{Assert Layout}, while rule-based evaluation enables automated interaction and basic verification, it remains limited in identifying global rendering deviations such as disproportionate scaling, overlapping elements, or misalignments.
To address this, we trained a specialized visual evaluation model utilizing a dual-stream architecture based on ResNet-50~\cite{he2016resnet}.
This model processes pairs of reference screenshots and generated screenshots as input to learn high-dimensional visual representations in the feature space.
It ultimately outputs a similarity score $s \in [0, 1]$, thereby quantifying the perceptual discrepancy between the rendered interface and the authentic target. 

Due to the scarcity of large-scale annotated data for GUI layout alignment, we designed an automated synthesis pipeline to generate training samples via procedural perturbations of real interfaces.
Specifically, we utilized Gemini-3-Flash to reconstruct 2048 GUI interfaces from our benchmark into high-fidelity HTML/CSS code representations.
To further enhance the model's generalization to real-world distributions, we additionally integrated 484 real-webpage instances from Design2Code~\cite{SiZLYLY2025Design2Code}.

For each original instance, ten variants were constructed to cover a diverse range of layout scenarios, ranging from minor fluctuations to complete structural failure.
First, three variants underwent random proportional scaling within a $\pm$30\% range; as scaling preserves the layout structure, their similarity labels were set to 1.0.
This subset encourages the model to focus on overall layout correctness rather than pixel-level differences.
Subsequently, by randomly injecting varying degrees of element deletion, positional shifts, layout collapse and style tampering at the HTML/ CSS code level, we constructed five additional variants with increasing layout damage.
During this process, we introduced a weight-based penalty mechanism that linearly deducts the initial score (1.0) based on the extent of structural damage, to assign precise labels to these variants.
The last two varants were randomly selected from other unrelated instances, with labels set to 0.
This construction strategy ensures that the model can learn a full spectrum of visual representations.

In total, we curated a dataset of 25k pairs, split 8:1:1 for training/validation/ testing.
Experiments show our model achieves a low MAE of 0.09 on the test set, proving its efficacy in detecting fine-grained layout discrepancies.
Consequently, this provides a robust visual assessment for GUI generation tasks.

 \begin{figure*}[t]
  \centering
  \includegraphics[width=\linewidth, trim=2.2cm 3.5cm 2.2cm 2.3cm, clip]{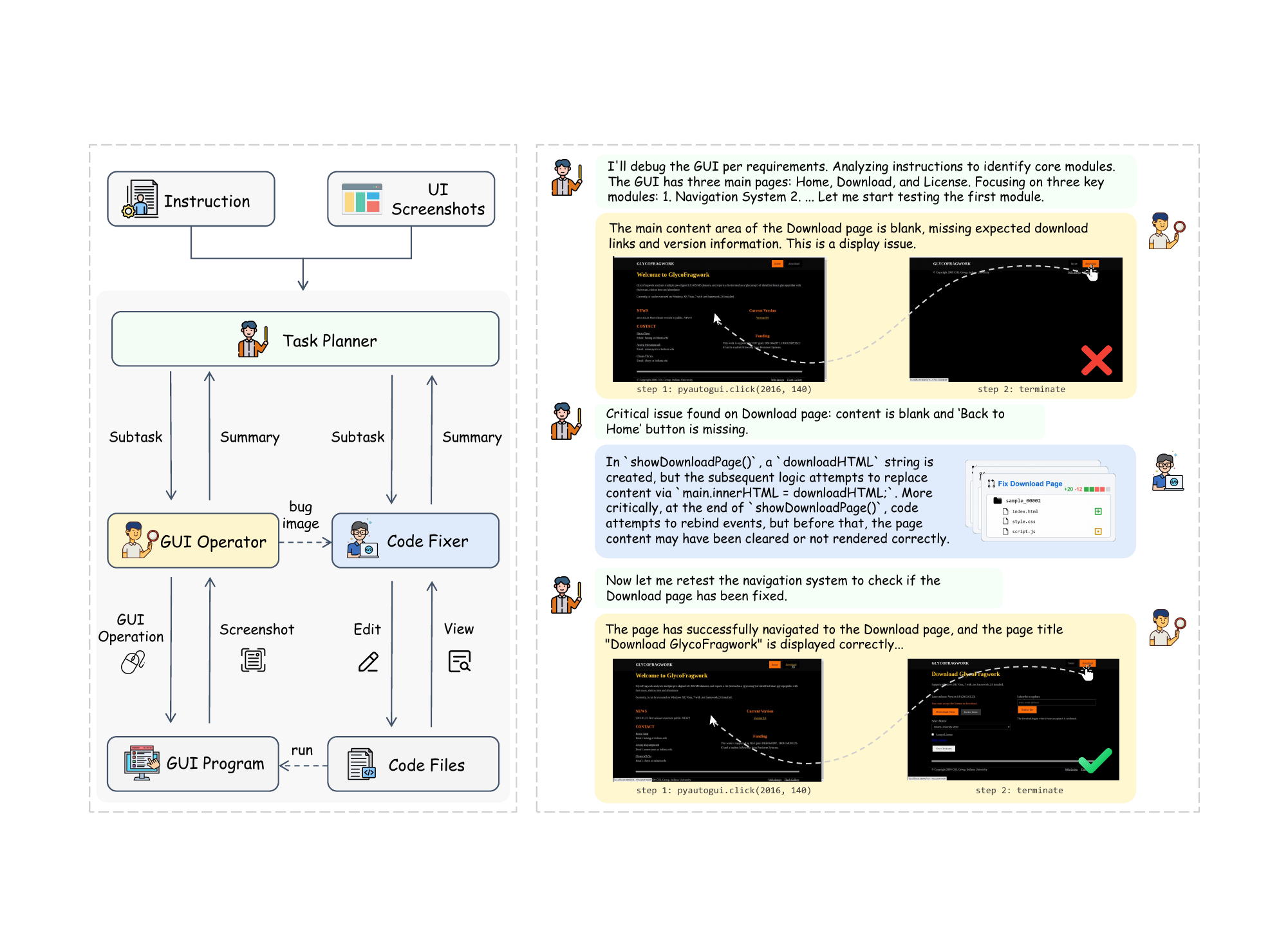}
    \caption{Overall architecture (left) and interaction details(right) of VF-Coder. Zoom in to see better. The right presents a real case of VF-Coder: The GUI Operator detects a rendering error on the download page through visual interaction. Under the coordination of the Task Planner, the Code Fixer successfully identifies and repairs the issue.}
  \label{vf_coder}
\end{figure*}

\section{VF-Coder}

Fig.~\ref{vf_coder} shows the overall framework of our proposed VF-Coder, which consists of three agents, including Task Planner, Code Fixer, and GUI Operator, that collaboratively analyze and repair logical and visual errors of GUI apps, effectively enhancing code quality.

\subsection{Task Planner}

The Task Planner serves as the central controller of VF-Coder, responsible for task decomposition, agent coordination, and overall workflow management. 
It consists of a MLLM equipped with a session memory module. 
Upon receiving task instruction and UI screenshots, the Task Planner divides the task into a sequence of subtasks and assigns them to the appropriate agents (\ie GUI Operator for runtime interaction and validation or Code Fixer for code modification). 

After each subtask, Task Planner collects the agents’ feedback, updates its internal state, and determines the next action accordingly. 
Once all detected issues are resolved, the Task Planner terminates the debugging process and produces the final GUI app.

\subsection{GUI Operator}

GUI Agents (or Computer-Use Agent, CUA)~\cite{song2025coact, yang2025gta1, sun2025guixplore} have emerged as a prominent research frontier in recent years. 
By transforming the user instruction into step-by-step executable operations on the screen, GUI Agents are revolutionizing the paradigm of human-computer interaction~\cite{nguyen2024guiagentsurvey}. 
Inspired by this paradigm, we design GUI Operator, a vision-capable GUI agent that autonomously navigates and interacts with app interfaces to identify runtime execution anomalies.

The GUI Operator receives the task instructions, screenshots, and specific inspection task from the Task Planner.
Upon invocation, the GUI Operator first initializes and launches the target app within an isolated sandbox environment. 
By perceiving real-time screenshots and terminal logs, the operator predicts and executes the corresponding interaction operations. 
The environment then automatically updates and returns the interface state, facilitating subsequent interactions within an iterative feedback loop.  
Through this process, the GUI Operator can proactively detect logical and visual errors during runtime.
Once an issue is detected, it immediately terminates the current session and sends feedback to the Task Planner for subsequent repair and revalidation steps.

\subsection{Code Fixer}

The Code Fixer is responsible for fixing the GUI code based on the task instruction, UI screenshots and the bug description provided by the Task Planner.
After being invoked, it automatically parses the source files, localizes the erroneous code segments, and synthesizes corresponding patches. 
After completing the repair, the Code Fixer summarizes the modifications to the Task Planner, which subsequently determines whether the app requires further validation or additional debugging iterations. 

\subsection{Memory and Communication Management}

Each agent maintains an independent session memory that records its local context within the current debugging session. 
After completing a subtask, both GUI Operator and Code Fixer send feedback to the \textit{Task Planner}, which integrates it into its global context for subsequent decision-making. 
To ensure task focus and prevent cross-interference between subtasks, temporary histories of GUI Operator and Code Fixer are cleared after each execution, while the Task Planner retains a persistent high-level memory of the overall workflow.

\subsection{Discuss with Kimi agent}
\label{dicuss-kimi}
We note that the Kimi agent recently introduced a feature for debugging GUI applications using screenshots. 
Consequently, a direct comparison or in-depth analysis is not feasible. 
To still provide context, we conduct a manual evaluation of its performance on our proposed InteractGUI Bench. 
The results of this evaluation are presented and discussed in the Experiment section.

\section{Experiments}
\label{sec:experiments}

\subsection{Experimental Setup}

\subsubsection{Models and Code Agent Frameworks:}
In this section, we first conduct a systematic evaluation of five mainstream MLLMs on InteractGUI Bench to establish performance baselines, including GPT-5.2~\cite{openai2025gpt5_2}, Claude-4.5-Sonnet~\cite{anthropic2025claude4_5}, Gemini-3.1-Pro~\cite{deepmind2025gemini_3_1_pro}, Gemini-3-Flash~\cite{deepmind2025gemini_3_flash} and Kimi-2.5~\cite{team2026kimi}.

Then, we use Gemini-3-Flash as the backbone model, comparing our VF-Coder with two representative frameworks:
1) Gemini CLI~\cite{gemini_cli_team2025gemini_cli}, an open-source AI agent developed by Google that performs autonomous planning, shell command execution, and web access;
2) Cursor CLI~\cite{cursor_team2025cursor}, the command-line version of Cursor that provides full access to Cursor’s features, including codebase indexing and automated code refinement via linter integration. 

\subsubsection{Implementation Details:}
All models and code agents are prompted to generate GUI code based on PySide6 (see Supplementary for detailed prompts).
We keep the default configurations for all Baseline models and compared frameworks.
For VF-Coder, to balance inference cost and performance, we set the maximum number of planning iterations for the Task Planner to 10. 
The GUI Operator is allowed to execute at most 5 steps for each subtask while retaining the most recent 4 interaction histories as context.
The framework automatically terminates if the step limit is reached. 

\subsubsection{Evaluation Metrics}
Following SWE-bench~\cite{jimenez2024swebench}, we adopt \textbf{\% Resolved} (pass@1), \textbf{\$ Avg. Cost} and \textbf{Avg. Visual Score} to evaluate task completion, efficiency, and visual fidelity: \% Resolved represents the percentage of tasks that pass all operations defined in the IES; \$ Avg. Cost denotes the average economic cost per task. Under the same model settings, this directly reflects token consumption and computational overhead; Avg. Visual Score measures the average visual similarity between rendered and ground-truth interfaces via our pre-trained visual evaluation model. 

To further identify performance bottlenecks, we additionally report four fine-grained metrics: \textbf{FS} (Fail-to-Start), \textbf{AE}, \textbf{AC}, and \textbf{CK}. FS is the ratio of tasks that fail to launch due to compilation or runtime errors; AE is the success rate of \textbf{Assert Element} operations, reflecting the presence of essential UI components; AC is the success rate of \textbf{Assert Color} operations, reflecting the accuracy of visual style; CK is the success rate of \textbf{Click} operations, reflecting the logical integrity of interactive functionalities.

\begin{table*}[t]
{
  \small
  \caption{Performance comparison of baseline LLMs under Desktop (Pyside6) settings on proposed InteractGUI Bench. 
           All metrics are reported in percentages except Avg. Cost and Avg. Visual Score. Best results are \textbf{bolded}.}
  \label{baseline_performance}
  \centering
  \resizebox{\textwidth}{!}{%
  \begin{tabular}{c|ccc|cccc}
    \toprule
        Model & \% Resolved ↑ |& \$ Avg. Cost ↓ | &Avg. Visual Score ↑& FS ↓ & AE ↑& AC ↑ & CK ↑ \\
    \midrule
        GPT-5.2 & \textbf{26.99}& 0.1746  &\textbf{0.4839}& \textbf{14.53}& \textbf{58.28}& \textbf{67.53}& \textbf{72.52}\\
        Claude-4.5-Sonnet & 23.18& 0.1563& 0.3378& 15.57& 53.94& 62.75& 66.50\\
 Gemini-3.1-Pro& 23.57& 0.2076& 0.4816& 21.66& 51.47& 57.36&61.03\\
        Gemini-3-Flash & 21.68& 0.0147 &0.4284& 25.20& 49.05& 58.35& 58.62\\
        Kimi-2.5 & 11.31& \textbf{0.0202}&0.3867& 27.08& 34.04& 45.86& 45.77\\   
    \bottomrule
  \end{tabular}
  }
}
\end{table*}

\subsection{Results on proposed InteractGUI Bench}

\subsubsection{Baseline Models Performance:}

Based on the experimental results in Table~\ref{baseline_performance} , we make the following observations:
1) As shown in the table, GPT-5.2 achieves the highest task success rate at 26.99\%, closely followed by Gemini-3.1-Pro. This indicates that even the top-performing models are still far from saturating the InteractGUI Bench, highlighting that our benchmark remains significantly challenging for current MLLMs.
2) In terms of visual similarity, GPT-5.2 also achieves the best performance with a score of 0.4839. There exists a loose correlation between visual scores and task success rates, as fully functional GUI applications are relatively less prone to severe visual discrepancies.

\begin{table}[t]
{
    \small
    \caption{Performance comparison of related code agent methods and VF-Coder under the Desktop (Pyside6) setting.}
    \label{comparation_with_existing_method}
    \centering
    \resizebox{\textwidth}{!}{%
        \begin{tabular}{l|ccc|cccc}  % X 列会自动分配宽度
        \toprule
            Method
            & \% Resolved ↑ | & \$ Avg. Cost ↓ | & Avg. Visual Score ↑ & FS ↓& AE ↑& AC ↑ & CK ↑ \\
        \midrule
            Gemini-3-Flash & 21.68& 0.0147 & 0.4284& 25.20& 49.05& 58.35& 58.62\\
            +Gemini CLI& 24.81& \textbf{0.1309}& 0.5190& 7.14& 50.69& 66.91& 61.66\\
            +Cursor CLI& 25.46& 0.1726& 0.5028& \textbf{1.59}& 53.45& 66.95& \textbf{73.64}\\
            \textbf{+VF-Coder} & \textbf{28.29}& 0.2064& \textbf{0.5584}& 4.47& \textbf{58.41}& \textbf{75.93}& 71.33\\
        \bottomrule
        \end{tabular}
    }
}
\end{table}

\subsubsection{VF-Coder vs. Text-Only Agents:}
\label{comparation_results}

As shown in Table~\ref{comparation_with_existing_method}, the baseline Gemini-3-Flash achieves a `\% Resolved' rate of 21.68\%.
With the integration of our VF-Coder framework, this metric increases to 28.29\%, representing an absolute improvement of 6.61\%.
This gain exceeds those achieved by text-based methods in the comparison.
In addition, VF-Coder also achieves the highest Avg. Visual Score (0.5584) and demonstrates consistent improvements across fine-grained interaction metrics, such as raising AE to 58.41\% and AC to 75.93\% .
These results indicate the effectiveness of VF-Coder in fixing visual problems.

In contrast, while Gemini CLI and Cursor CLI decrease the `FS'  to 7.14\% and 1.59\% respectively, their impact on visual integrity is limited.
This suggests that although text-based CLI feedback helps resolve execution-level errors, it lacks the visual information required to maintain layout consistency.
This demonstrates that traditional text-based agent frameworks have inherent limitations in GUI tasks, whereas visual feedback provides substantial informational gain for GUI debugging and repair.

\begin{table}[t]
\centering
\caption{Performance comparison of Kimi Agent and VF-Coder under the Desktop (PySide6) setting.
Some metrics are not reported due to the absence of a public API.}
\label{comparation_with_kimi_agent}
\resizebox{0.8\linewidth}{!}{%
    \begin{tabular}{l|cc|cccc}
    \toprule
        Method& \% Resolved ↑ & Avg. Visual Score ↑ & FS ↓ & AE ↑ & AC ↑ & CK ↑ \\
    \midrule
        Kimi-K2.5 & 16.25 & 0.4865 & 10.00 & 43.67 & 44.73 & 42.28\\
        +Kimi Agent & 18.75 & \textbf{0.5648} & \textbf{1.25} & \textbf{54.28} & 61.04 & 52.31\\
        \textbf{+VF-Coder} & \textbf{25.00} & 0.5530 & 3.75 & 52.50 & \textbf{66.04} & \textbf{61.66}\\
    \bottomrule
    \end{tabular}%
}
\end{table}

\subsubsection{Comparison with Kimi agent:} 
As discussed in Sec.~\ref{dicuss-kimi}, we include a comparison with the Kimi agent to provide additional context. 
Due to the absence of a public API for the Kimi agent, direct programmatic evaluation was not feasible. Instead, we conducted a manual evaluation by uploading the screenshots and prompts from a subset of 80 applications from our InteractGUI Bench to its web interface and recording the responses.

The results of this manual comparison are presented in Tab.~\ref{comparation_with_kimi_agent}. Our proposed VF-Coder achieves a much higher `\% Resolved' score (25.00 vs. 18.75) on this subset, showing the strong effectiveness of our approach.

\subsection{Ablation Study}

To further analyze the role of visual feedback, we decompose the GUI code generation task into the debugging and fixing stages.
All ablation experiments are conducted on the InteractGUI Bench using Gemini-3-Flash as the base model. 
In the debugging stage, we investigate the impact of visual interaction by enabling or disabling the GUI Operator within VF-Coder (denoted as w/ or w/o GUI Operator). When the GUI Operator is disabled (w/o GUI Operator), the system relies exclusively on textual interactions, such as inspecting source code or executing shell commands, to perform debugging tasks.
In the fixing stage, we examine the effect of visual feedback by controlling whether the Code Fixer receives the GUI screenshot captured at the moment the bug is detected (w/ or w/o bug screenshot). 
Without the screenshot (\textit{w/o bug screenshot}), the Code Fixer lacks direct visual access to the interface state at the time of the error. Instead, it must rely solely on the textual bug descriptions provided by the GUI Operator to locate and repair the faulty code.

As shown in Table~\ref{ablation}, the contribution of visual feedback varies significantly across stages.
In the debugging stage, incorporating visual information (w/o bug screenshot) increases the task resolution rate (\%Resolved) from 24.79\% under text-only interactions (w/o GUI Operator) to 27.64\%, an improvement of 2.85\%, while the average visual quality score (Avg. Visual Score) rises from 0.5193 to 0.5390.
This indicates that visual feedback plays a crucial role in error detection and interface state understanding, and can effectively improve the visual quality of generated interfaces.
In the fixing stage, further introducing visual feedback (Full) yields only marginal gains in task resolution rate, with a mere 0.65\% increase (27.64\%→28.29\%).
However, there remains improvement in visual fidelity metrics (Avg. Visual Score increases from 0.5390 to 0.5584), suggesting that bug screenshots are helpful in conveying visual errors but have limited utility for other types of errors such as functional logic.

\begin{table*}[t]
  \caption{Ablation study on the effects of different agents and visual feedback.}
  \label{ablation}
  \centering
  \resizebox{\textwidth}{!}{
  \begin{tabular}{l|cc|ccc|cccc}
    \toprule
        & Debugging& Fixing
        & \% Resolved ↑ |& \$ Avg. Cost ↓ |& Avg. Visual Score ↑& FS ↓& AE ↑& AC ↑ & CK ↑ \\
    \midrule
    w/o GUI Operator   
        & textual & textual 
        & 24.79& \textbf{0.1351}& 0.5193& 3.25& 50.69& 61.91& 59.63\\
        
    w/o bug screenshot 
        & visual & textual 
        & 27.64& 0.1963& 0.5390& 5.28& 53.45& 66.95& \textbf{73.64}\\ 
        
    Full   
        & visual & visual
        & \textbf{28.29}& 0.2064& \textbf{0.5584}& \textbf{4.47}& \textbf{58.41}& \textbf{75.93}& 71.33\\
    \bottomrule
  \end{tabular}
  }
\end{table*}

\begin{figure*}[t]
  \centering
  \includegraphics[width=\linewidth, trim=2cm 0.5cm 2cm 0.5cm, clip]{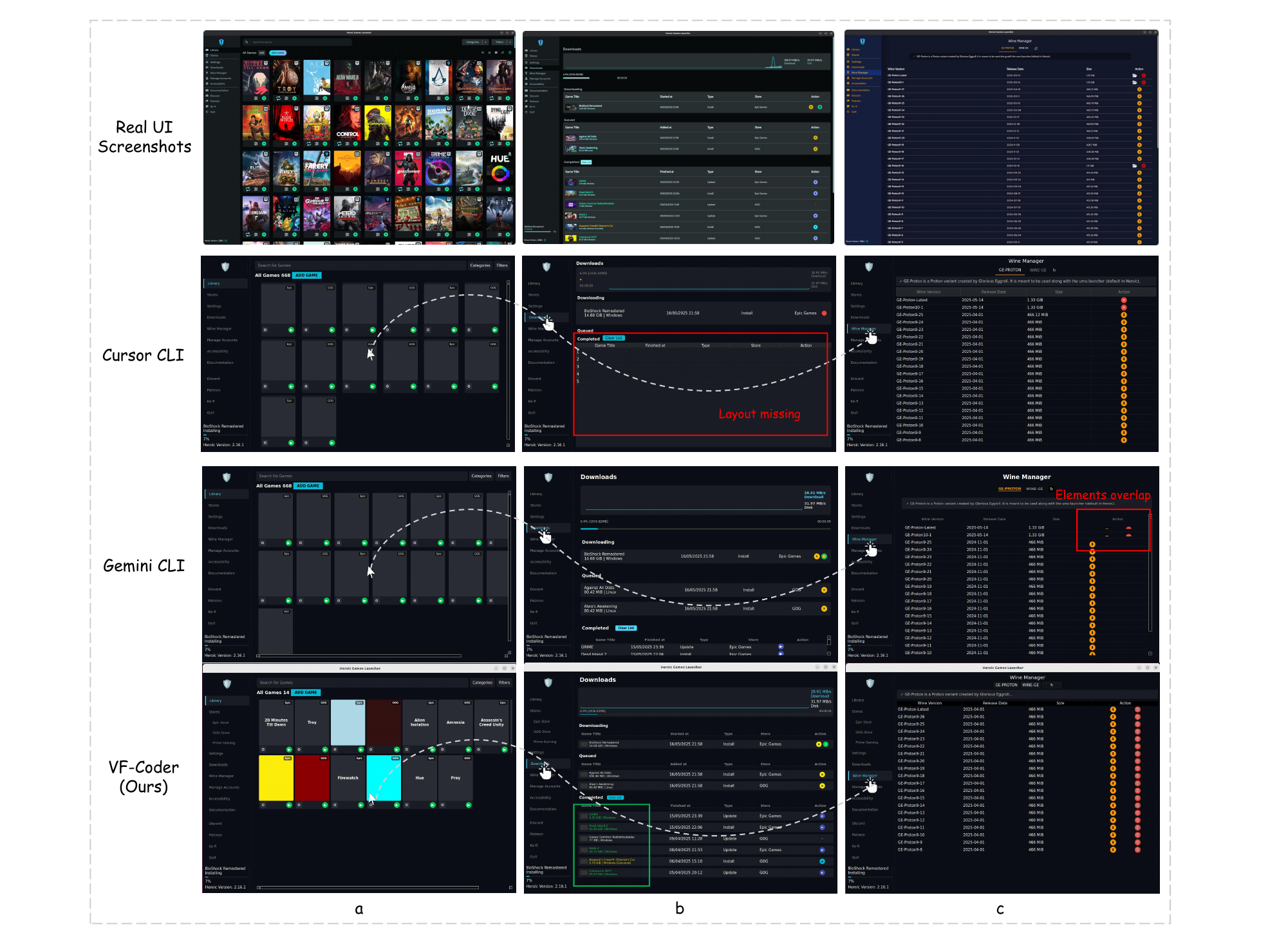}
    \caption{Visual comparison of Cursor CLI, Gemini CLI and VF-Coder in a real case.}
  \label{case_study}
\end{figure*}

\subsection{Case Study}
\label{case_study}
To concretely illustrate the advantages and working mechanism of VF-Coder, we visualize the evaluation trajectories of Cursor CLI, Gemini CLI, and VF-Coder on a representative task in Figure~\ref{case_study}.
Due to the page limit, more comparison cases are presented in the appendix.
Our key observations are as follows: 
1) VF-Coder achieves the closest visual fidelity to the Real UI Screenshots. Cursor CLI exhibits significant layout deviations in interface b, while Gemini CLI, suffers from element overlapping in interface c. These are difficult-identified problems for text-output-based methods. With the support of visual perception, VF-Coder not only avoids these obvious visual errors , but also attempts to align with fine-grained style details, such as the precise color matching of fonts in interface b (highlighted by green boxes in the figure), which shows the advantage of Visual Feedback. 2) Despite VF-Coder achieving the best performance, all methods still exhibit visible gaps compared to the Real UI Screenshots. This limitation likely stems from the inherent capability boundaries of base models, reflecting the common challenges faced by current approaches in generating real-world desktop GUI applications.

\section{Conclusion}
\label{sec:conclusion}

Existing code agents primarily rely on textual feedback to ensure the quality of generated code, which presents significant limitations in GUI code generation scenarios. To systematically evaluate this challenge, we propose InteractGUI Bench, the first benchmark that assesses GUI code generation through interactive execution and visual verification. By simulating realistic user behaviors and jointly examining functional logic and visual structure, this benchmark provides a comprehensive measurement of model capability in GUI contexts. Building upon it, we further introduce VF-Coder, a vision-feedback-based multi-agent framework that integrates interface perception and interactive exploration directly into the debugging loop. Extensive experiments show that VF-Coder significantly improves the reliability of GUI code over text-only agents, highlighting the crucial role of visual feedback in understanding interface states and repairing logical or rendering defects. We hope this study lays a foundation for future research on GUI code generation and visually grounded software intelligence.

\clearpage

\bibliographystyle{splncs04}
\bibliography{main}

\clearpage

\appendix

\setcounter{page}{1}
% \maketitlesupplementary

\section{More discussions of Proposed InteractGUI Bench}

In this section, we provide the statistical information of our InteractGUI Bench and compare it with other relevant benchmarks. We also present three real tasks from InteractGUI Bench in Figure~\ref{data_example_1},~\ref{data_example_2}, \ref{data_example_3}. 

\subsection{Data Statistics and Diversity}
\begin{figure}
    \centering
    \includegraphics[width=\linewidth, trim=4cm 2.3cm 2.5cm 3.5cm, clip]{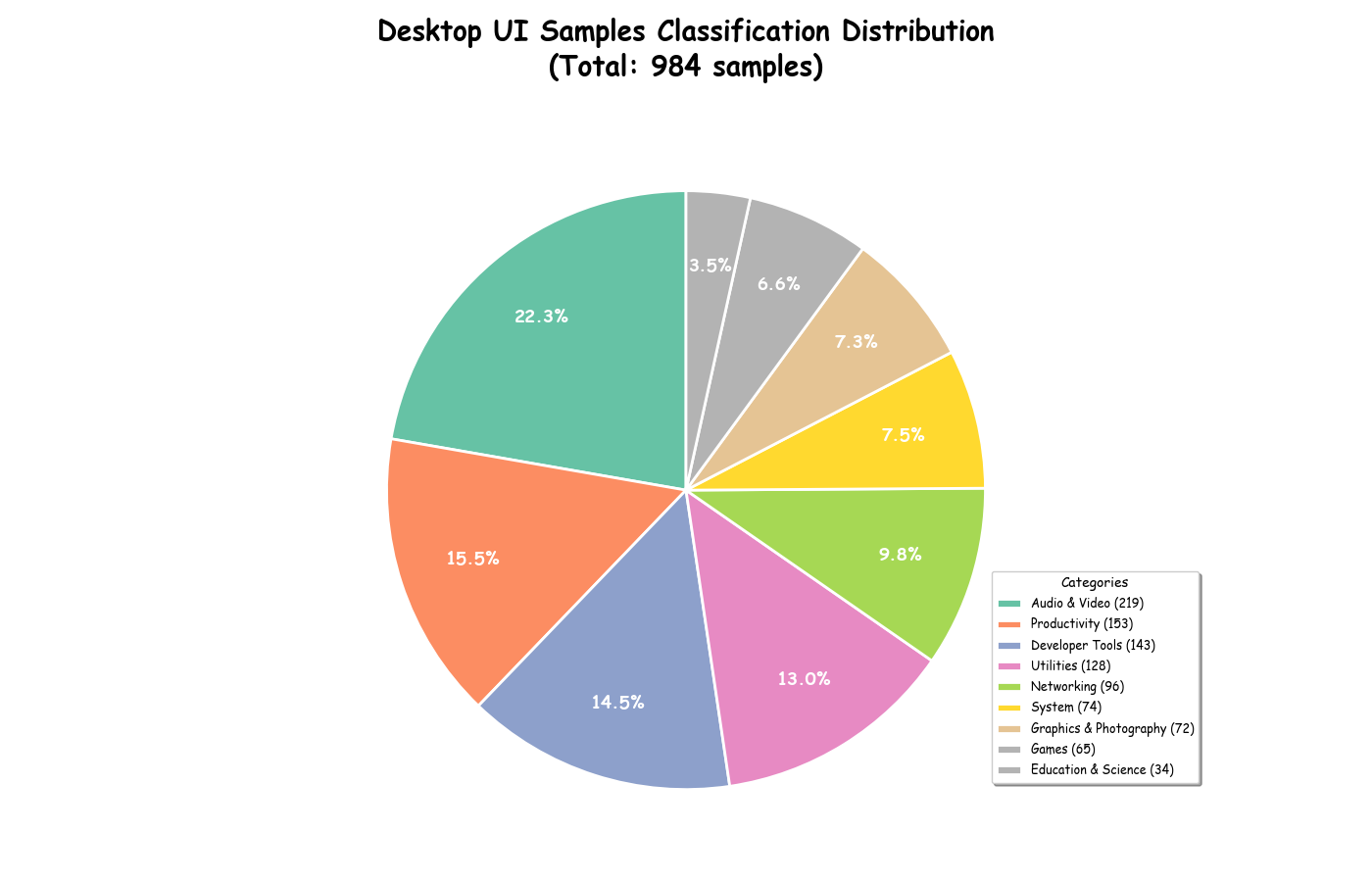}
    \caption{Task Category Composition of InteractGUI Bench}
    \label{classification_chart}
\end{figure}

We present several quantitative statistics to characterize the composition of our benchmark:
\begin{itemize}
    \item \textbf{Number of GUI Screenshots}: Our benchmark comprises 984 distinct desktop GUI application tasks, covering a total of 2,027 GUI screenshots. 
    On average, each task includes 2.06 screenshots, with a standard deviation of 1.45 and a range from 1 to 10.
    \item \textbf{Number of Evaluation Steps}: Following the definition in IES, we report the number of evaluation steps for each task. The average number of steps per task is 8.87, with a standard deviation of 5.09 and a range from 1 to 28.
    \item \textbf{Domain Distribution}: To illustrate the range of application domains covered, we categorize all tasks according to the 10 domain labels defined by Flathub. Figure~\ref{classification_chart} presents the resulting distribution as a pie chart. The most frequently represented categories are Audio \& Video, Productivity, and Developer Tools, followed by Utilities, Networking, System, Graphics \& Photography, Games, and Education \& Science. 
\end{itemize}

\subsection{Comparison with prior benchmarks}
We compare our proposed InteractGUI Bench with existing relevant benchmarks in Table~\ref{comparation_with_previous_work}.
Overall, the proposed InteractGUI Bench is the first to achieve both high granularity (GR) and strong human alignment (CHA) in desktop scenarios.

Here, GR refers to fine-grained evaluation spanning from individual components to overall layout, while CHA is reflected in our evaluation approach—instead of relying on MLLM-as-Judge as in related work, we adopt rule-based evaluation and visual evaluation model trained on in-domain data.

Additionally, it is worth noting that while both WebGen-Bench\cite{lu2025webgen} and our proposed VF-Coder method utilize GUI Agents, WebGen-Bench employs them solely for executing pre-defined test instructions to evaluate generated results, and does not explore visual-feedback-based code debugging strategies, which is a fundamental difference from our work.

% 定义颜色（根据图片中的颜色调整）
\definecolor{green}{RGB}{0, 128, 0} 
\definecolor{orange}{RGB}{255, 140, 0} 
\definecolor{gray}{RGB}{100, 100, 100} 
\begin{table}[h]
\centering
\vspace{2mm}
\caption{Comparison with prior benchmarks. \textbf{GR} (Granularity): the level of evaluation detail, ranging from individual UI components to overall interface layout. Higher is better (more comprehensive). \textbf{CHA} (Strong Human Alignment): the degree to which automated evaluation matches human judgment and preferences. Higher is better (more consistent with humans)}
\resizebox{\linewidth}{!}{
\begin{tabular}{l|cc|cccc}
\toprule
Benchmark& Size& Input Type& Eval Scope& Eval Method& GR↑& CHA↑\\
\midrule
HumanEval\cite{chen2021humaneval} & 164 & text & Code & Rule-based & \textcolor{gray}{Low} & \textcolor{green}{High} \\
Design2Code\cite{SiZLYLY2025Design2Code} & 484 & image & WebUI & Model-based & \textcolor{green}{High} & \textcolor{orange}{Mid} \\
FrontendBench\cite{zhu2025frontendbench} & 148 & text & WebUI & Rule-based & \textcolor{orange}{Mid} & \textcolor{green}{High} \\
WebGen-Bench\cite{lu2025webgen} & 101 & text & WebUI & Model-based & \textcolor{orange}{Mid} & \textcolor{orange}{Mid} \\
\midrule
\textbf{InteractGUI Bench} & 984 & image+text & DesktopUI &\quad Rule-based + Model-based \quad & \textcolor{green}{High} & \textcolor{green}{High} \\
\bottomrule
\end{tabular}
}
\label{comparation_with_previous_work}
\end{table}

\section{Additional Experiments}
We report additional experimental results of the proposed VF-Coder on two extra benchmarks: Design2Code-Hard\cite{SiZLYLY2025Design2Code} and WebGen-Bench\cite{lu2025webgen}. 

\subsection{Design2Code-Hard}
Design2Code is an open-source, high-quality dataset containing 484 real-world web pages, primarily used to evaluate Multimodal Large Language Models (MLLMs) in converting visual designs to HTML/CSS code.
We conduct experiments on its more challenging subset, Design2Code-Hard, which comprises 80 difficult examples with unique challenges (\eg 26\% of examples contain more than 500 HTML tags, and 19\% contain non-English content).
We compare VF-Coder with three baseline prompting methods in Design2Code-Hard: ``Direct'', ``Text-Augmented'', and ``Self-Revision''. All methods use Gemini-3-Flash\cite{deepmind2025gemini_3_flash} as the base model.

Table~\ref{design2code_exp} presents the performance comparison of different methods on Design-2Code-Hard.
VF-Coder achieves optimal performance in two critical visual dimensions: Position and Color, with scores of 88.15\% and 89.09\% respectively, validating the unique advantages of the visual feedback mechanism in enhancing layout precision and visual style fidelity.
In dimensions more reliant on textual feedback, namely Block and Text, VF-Coder attains accuracies of 77.83\% and 96.31\%, slightly underperforming Self-Revision.
This may be attributed to the text augmentation strategy employed by Self-Revision: it first extracts all text elements from the original webpage and appends them to the instruction prompt along with the screenshot input, thereby obtaining more comprehensive textual content information. 

\begin{table}
    \centering
    \vspace{4mm}
    \caption{Performance comparison on Design2Code-Hard}
    \begin{tabular}{l|cccc|c}
    \toprule
         Method&  Block ↑&  Text ↑&  Position ↑&  Color ↑& CLIP ↑\\
    \midrule
         Direct
&  76.04&  96.43&  88.04&  87.78& 90.64\\
         Text-Augment
&  73.47&  93.70&  82.89&  83.20& 89.89\\
         Self-Revision&  \textbf{78.09}&  \textbf{98.65}&  87.53&  87.85& \textbf{91.28}\\
         VF-Coder (Ours)&  77.83&  96.31&  \textbf{88.15}&  \textbf{89.09}& 91.07\\
    \bottomrule   
    \end{tabular}
    \label{design2code_exp}
\end{table}
\subsection{WebGen-Bench}
WebGen-Bench is a novel benchmark designed to evaluate code agents' ability to convert natural language instructions into multi-file website codebases from scratch.
The benchmark consists of 101 tasks and 647 test cases, employing an MLLM-as-judge paradigm for automatic evaluation.
In our experiments, Gemini-3-Flash serves as the judge model.
Since VF-Coder is not designed for building multi-file website codebases, we utilize VF-Coder to perform complex visual debugging and code repair based on the complete codebases generated by Bolt.diy, and compare with it.
All methods also use Gemini-3-Flash as the base model.

Table~\ref{webgen_exp} presents the performance comparison on WebGen-Bench.
VF-Coder achieves substantial improvements over Bolt.diy, with accuracy increasing from 31.8\% to 41.8\% and appearance score rising from 2.74 to 3.96, demonstrating the effectiveness of visual feedback in identifying and rectifying rendering errors in complex multi-file codebases.
Table~\ref{webgen_category_results} further breaks down the results by category.
VF-Coder consistently outperforms Bolt.diy across all six categories, with particularly pronounced gains in Content Presentation (+19.2\%) and Data Display Testing (+14.2\%).
These categories heavily rely on precise visual layout, where VF-Coder's iterative visual debugging mechanism proves especially beneficial.
Overall, the results demonstrate the general effectiveness of visual feedback in improving code generation quality, across both desktop GUI apps and complex website codebases.

% We report results following the official metrics specified by Design2Code and WebGen-Bench; please refer to their main paper for detailed specifications. 

\begin{table}
    \centering
    \vspace{4mm}
    \caption{Performance comparison on WebGen-Bench}
    \begin{tabular}{l|cccc|cc}
    \toprule
         Method&  Yes &  Partial&  No &  Start Failed ↓& Accuracy ↑ &Appearance Score ↑\\
    \midrule
         Bolt.diy&  28.6&  6.3&  50.4&  14.7& 31.8&2.74\\
         VF-Coder (Ours)&  38.5&  6.6&  40.2&  14.7& \textbf{41.8}& 3.96\\
    \bottomrule   
    \end{tabular}
    \label{webgen_exp}
    \vspace{4mm}
\end{table}

\begin{table}
\centering
\caption{Category-wise evaluation results. The first three columns represent categories of website-generation instructions, while the last three represent categories of test cases. The highest score in each category is marked in bold.}
\resizebox{\linewidth}{!}{
\begin{tabular}{lcccccc}
\toprule
\multirow{2}{*}{Method}& \multicolumn{3}{c}{Instruction Categories} & \multicolumn{3}{c}{Test Case Categories} \\
\cmidrule(lr){2-4} \cmidrule(lr){5-7}
& Content& User& Data& Functional& Data& Design\\
& Presentation& Interaction& Management& Testing& Display& Validation\\
& & & & & Testing& Testing\\
\midrule
Bolt.diy & 25.9 & 32.7 & 36.2 & 24.5 & 36.6 & 44.7 \\
VF-Coder (Ours) & \textbf{45.1} & \textbf{38.3} & \textbf{45.0} & \textbf{30.8} & \textbf{50.8} & \textbf{58.6} \\
\bottomrule
\end{tabular}
}
\label{webgen_category_results}
\end{table}

\section{Visualization Analysis}
\label{sec:visualization}

\begin{figure}
    \centering
    \includegraphics[width=1\linewidth, trim=0cm 9cm 0cm 10cm, clip]{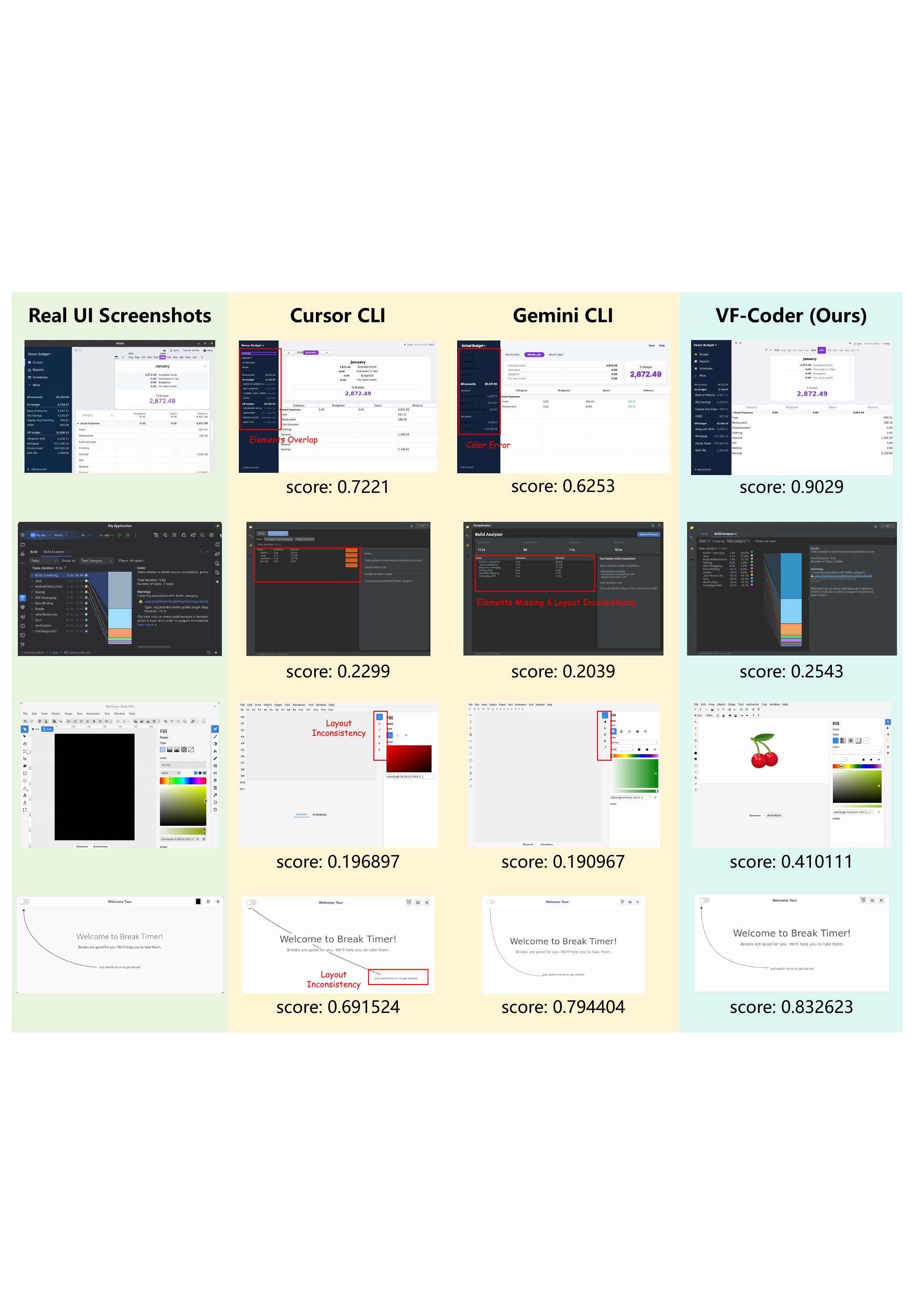}
    \vspace{-6mm}
    \caption{Visual comparison of rendered interfaces by Cursor CLI, Gemini CLI, and VF-Coder on representative InteractGUI Bench tasks. The Visual Score is computed by our Visual Evaluation Model. In these cases, our VF-Coder achieves the highest visual fidelity to real GUI screenshots and the highest visual score}
    \label{case_with_comparition}
\end{figure}

To further demonstrate the effectiveness of the proposed VF-Coder, we compare the rendered interfaces of Cursor CLI, Gemini CLI, and VF-Coder on several representative tasks in Figure~\ref{case_with_comparition}.

As shown in these examples, VF-Coder avoids obvious visual errors (\eg layout inconsistency, missing elements) and achieves rendering results that most closely resemble the real GUI screenshots, which demonstrates the advantage and necessity of visual feedback.
However, as discussed in the main paper (section 5.4), there remain visible discrepancies between the rendered interfaces and the real GUI screenshots, indicating that existing models still lack sufficient code generation capabilities and fine-grained visual perception abilities.

\section{Details of Visual Evaluation Model}
\label{sec:visual_evaluation}

In this section, we present the training details of our visual evaluation model for model-based evaluation on InteractGUI Bench, and demonstrate its scoring accuracy through several visualization examples.

\subsection{Training Details}

We employ a Siamese Network\cite{koch2015siamese} architecture based on ResNet50 as our visual evaluation model.
The model adopts a multi-scale feature extraction strategy, combining global-scale ($1\times1$) and local-scale ($8\times8$) feature representations to capture overall page layout information and detailed control arrangement information, respectively.
These features are ultimately fused into a 512-dimensional feature vector for similarity computation.

The model is trained on our constructed 25K GUI layout dataset, which is partitioned into training, validation, and test sets with a ratio of 8:1:1.
The specific training configurations are as follows: input images are resized to $640\times640$ pixels using aspect-ratio-preserving scaling with black padding to square format; the batch size is set to 16; the optimizer adopts AdamW with an initial learning rate of $1\times10^{-4}$ and a weight decay coefficient of $1\times10^{-4}$; the total training epoch is 30.
For the loss function, we utilize Mean Squared Error (MSE) Loss as the optimization objective.
We use Mean Absolute Error (MAE) as the evaluation metric, calculated as $\text{MAE} = \frac{1}{N} \sum_{i=1}^{N} |y_i - \hat{y}_i|$, where $y_i$ denotes the ground-truth similarity score, $\hat{y}_i$ denotes the predicted score, and $N$ represents the number of samples.

\begin{figure}
    \centering
    \includegraphics[width=1\linewidth, trim=0cm 0cm 0cm 0.5cm, clip]{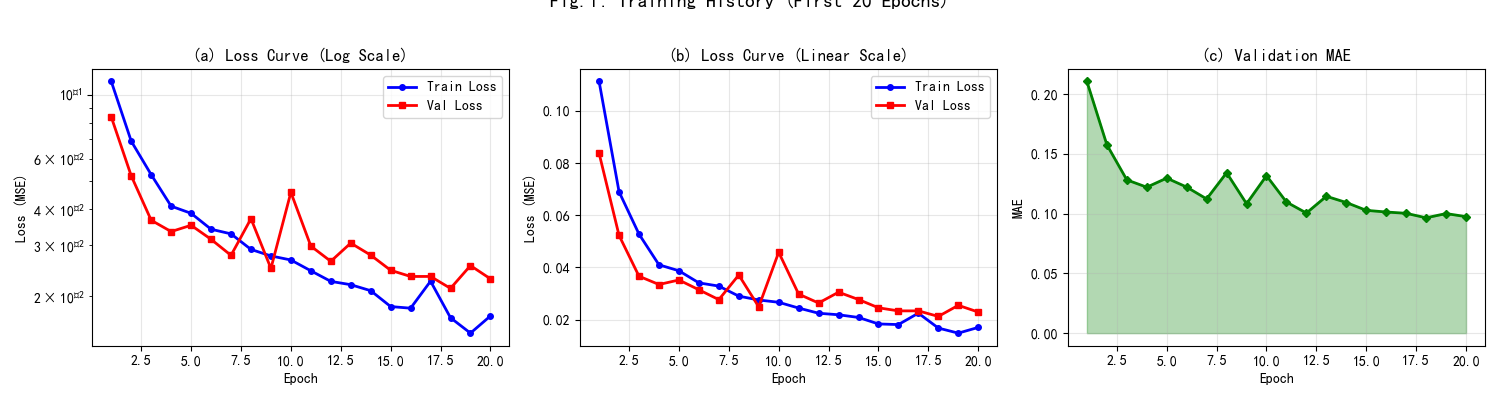}
    \caption{Training and validation MAE and loss curves of the Visual Evaluation Model}
    \label{training_log}
\end{figure}

Figure~\ref{training_log} illustrates the MAE curve on the training set and validation set during the training process.
Since the 18th epoch achieved the best MAE performance (0.0964) on the validation set, we ultimately select the model from the 18th epoch as our final visual evaluation model.

\subsection{Visualization}

\begin{figure}
    \centering
    \includegraphics[width=1\linewidth, trim=1.8cm 1.8cm 1.8cm 0.5cm, clip]{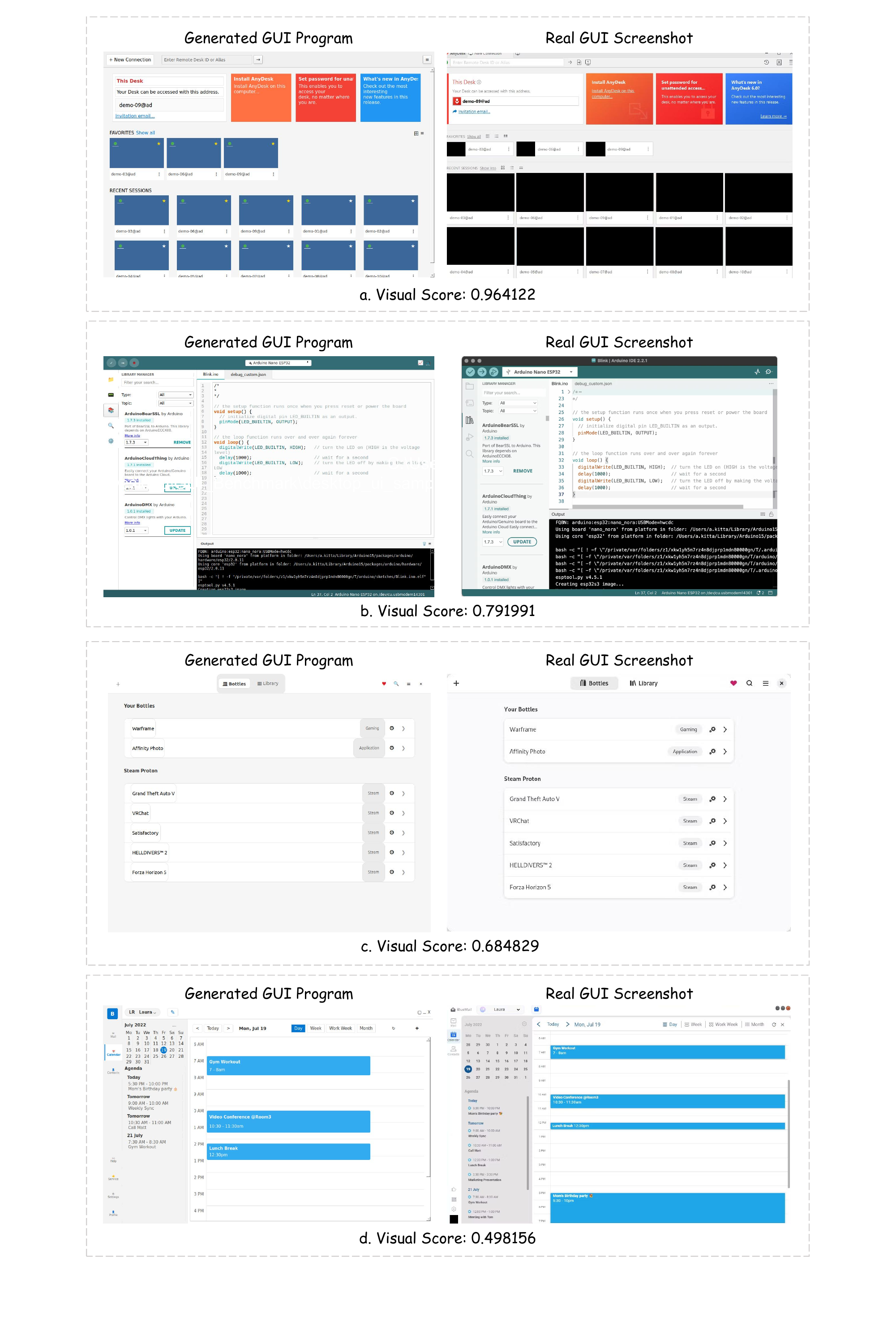}
    \caption{Qualitative visualization of Cases a–d ranked by Visual Evaluation Model scores (high to moderate). Scores correlate well with perceived visual similarity to real screenshots. }
    \label{visual_eval_case1}
\end{figure}

\begin{figure}
    \centering
    \includegraphics[width=1\linewidth, trim=1.8cm 17.3cm 1.8cm 1cm, clip]{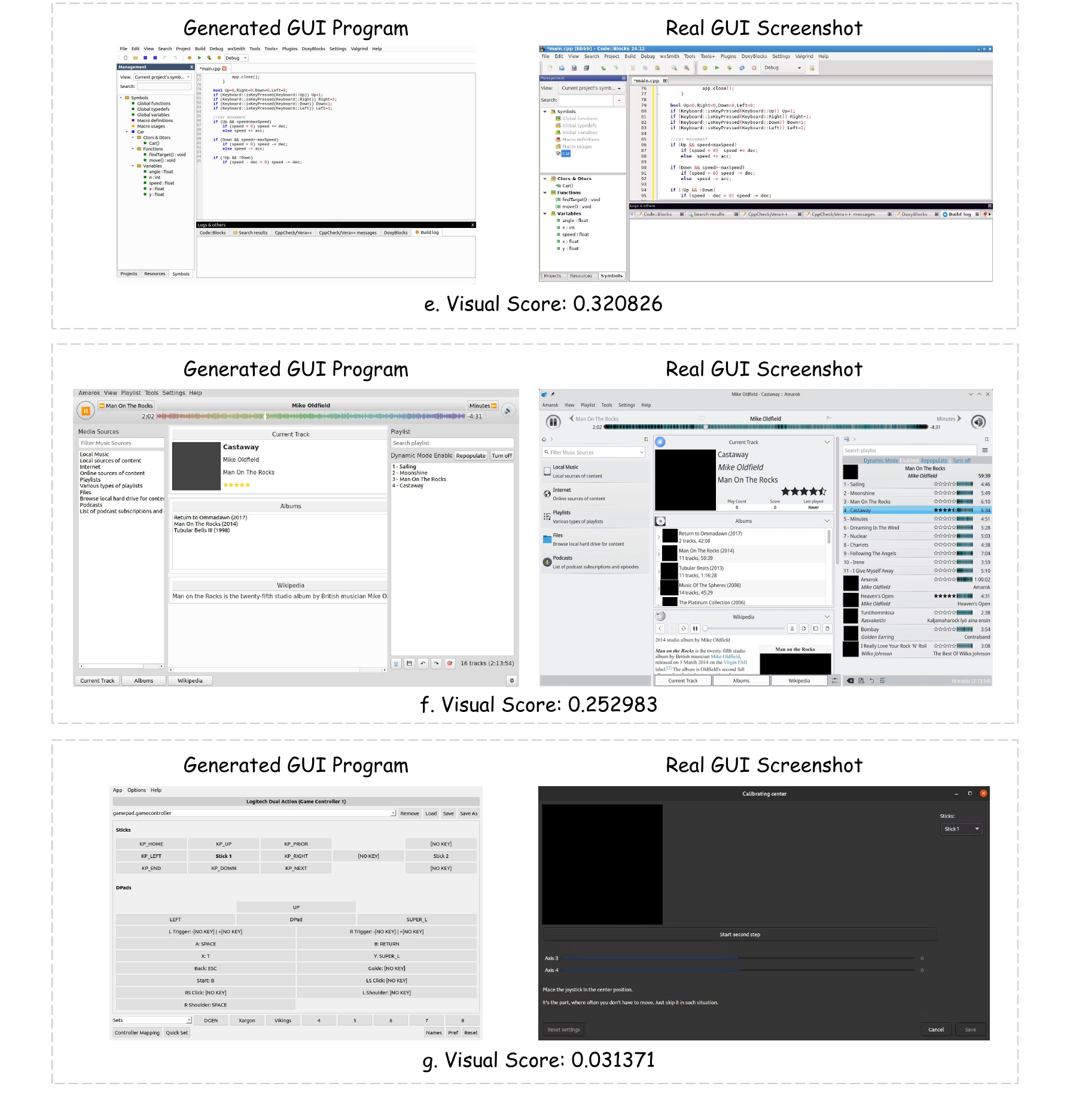}
    \caption{Qualitative visualization of Cases e–g ranked by Visual Evaluation Model scores (moderate to near-zero). Near-zero scores correctly indicate either significant rendering errors or fundamental differences in application layout styles} 
    \label{visual_eval_case2}
\end{figure}

We select a few real cases ranked by their visual evaluation model scores from high to low, as shown in Figures~\ref{visual_eval_case1} and \ref{visual_eval_case2}.
Through observation, we can find that the model scores generally align with human preferences, exhibiting a clear correlation between the scores and the similarity between reference images and generated images.
Meanwhile, interfaces from two completely different applications obtain scores close to 0, which is reasonable because layout styles of different applications should exhibit significant differences.

\begin{figure}
    \centering
    \includegraphics[width=1\linewidth, trim=0cm 3.3cm 0cm 2cm, clip]{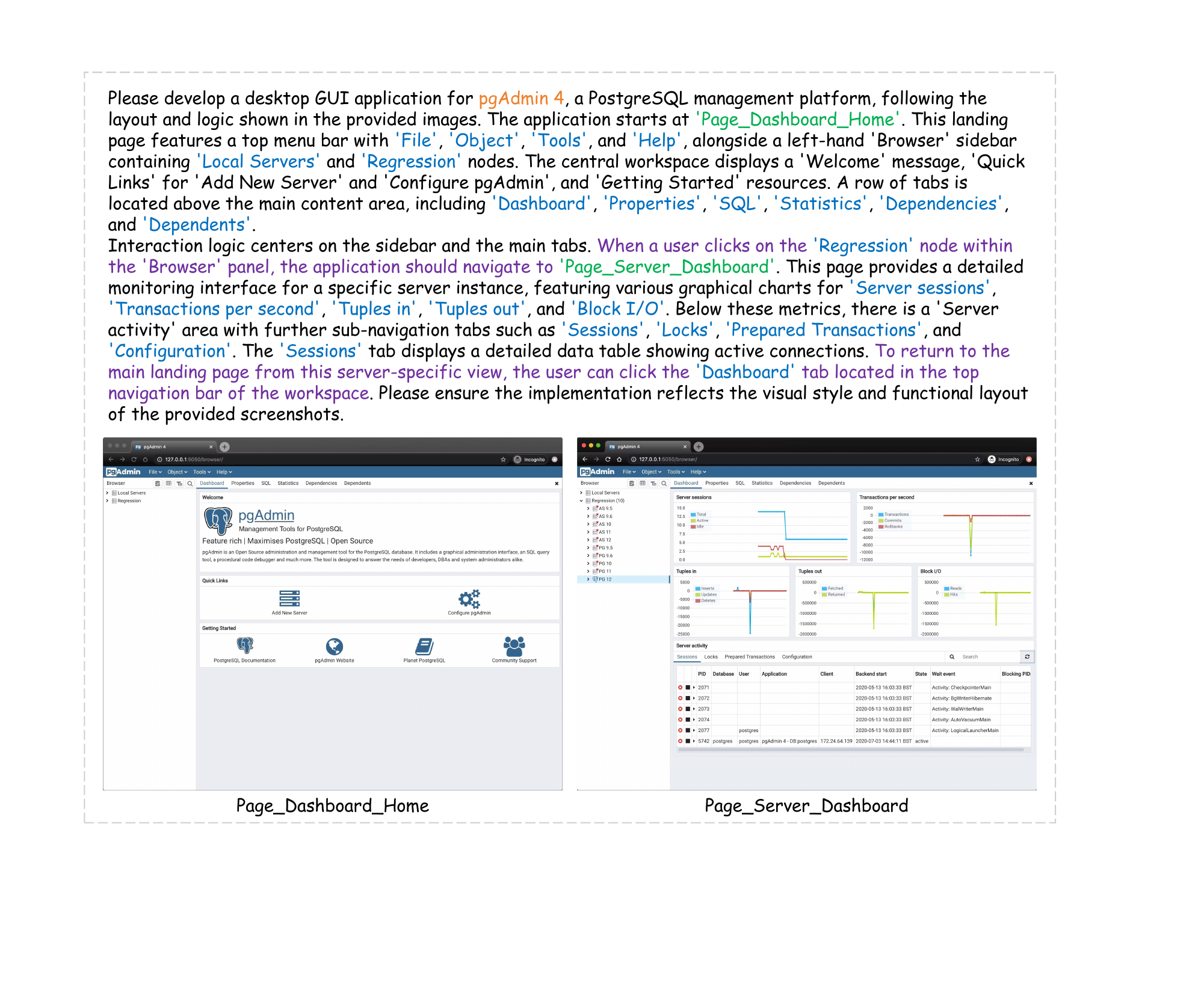}
    \caption{InteractGUI Bench Task Showcase}
    \label{data_example_1}
\end{figure}

\begin{figure}
    \centering
    \includegraphics[width=1\linewidth, trim=1cm 2.3cm 1cm 2cm, clip]{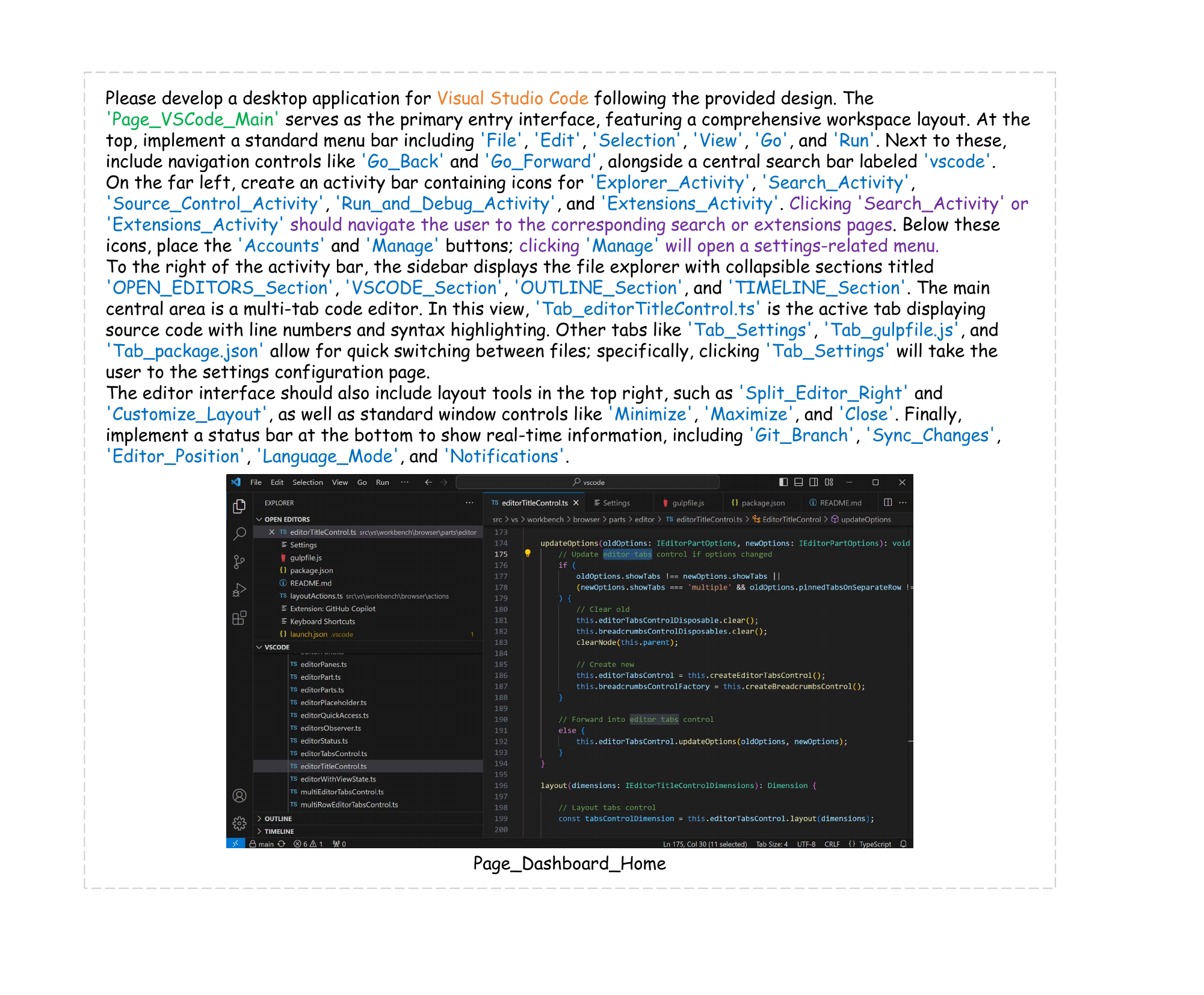}
    \caption{InteractGUI Bench Task Showcase}
    \label{data_example_2}
\end{figure}

\begin{figure}
    \centering
    \includegraphics[width=1\linewidth, trim=1cm 6.3cm 1cm 2cm, clip]{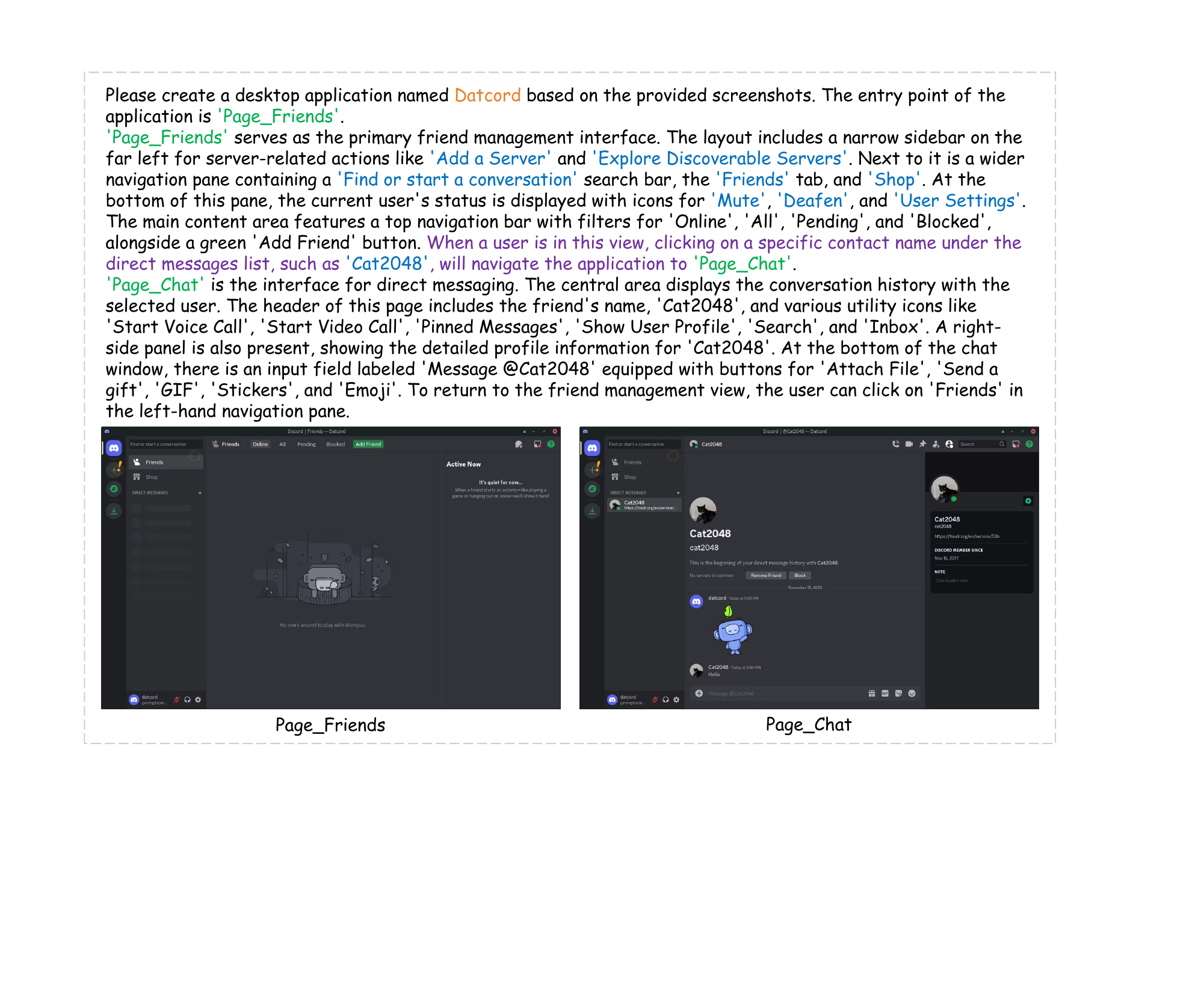}
    \caption{InteractGUI Bench Task Showcase}
    \label{data_example_3}
\end{figure}

\section{Further Details}

\subsection{Benchmark Task Examples}
We present three real tasks from InteractGUI Bench in Figure~\ref{data_example_1},~\ref{data_example_2}, \ref{data_example_3}. 
These GUI screenshots are sourced from three distinct real-world desktop applications: pgAdmin (a popular open-source database administration tool for PostgreSQL), VSCode (a widely-used code editor with an extensive plugin ecosystem), and Datcord (a Discord-style chat application).
In InteractGUI Bench, MLLMs or code agents are required to reconstruct the visual layouts and partial interaction logic of these real desktop-level applications, evaluating their code generation and GUI understanding capabilities in real-world scenarios.

\subsection{Prompt Details}
\begin{figure}
    \centering
    \includegraphics[width=1\linewidth, trim=1cm 29cm 1cm 1cm, clip]{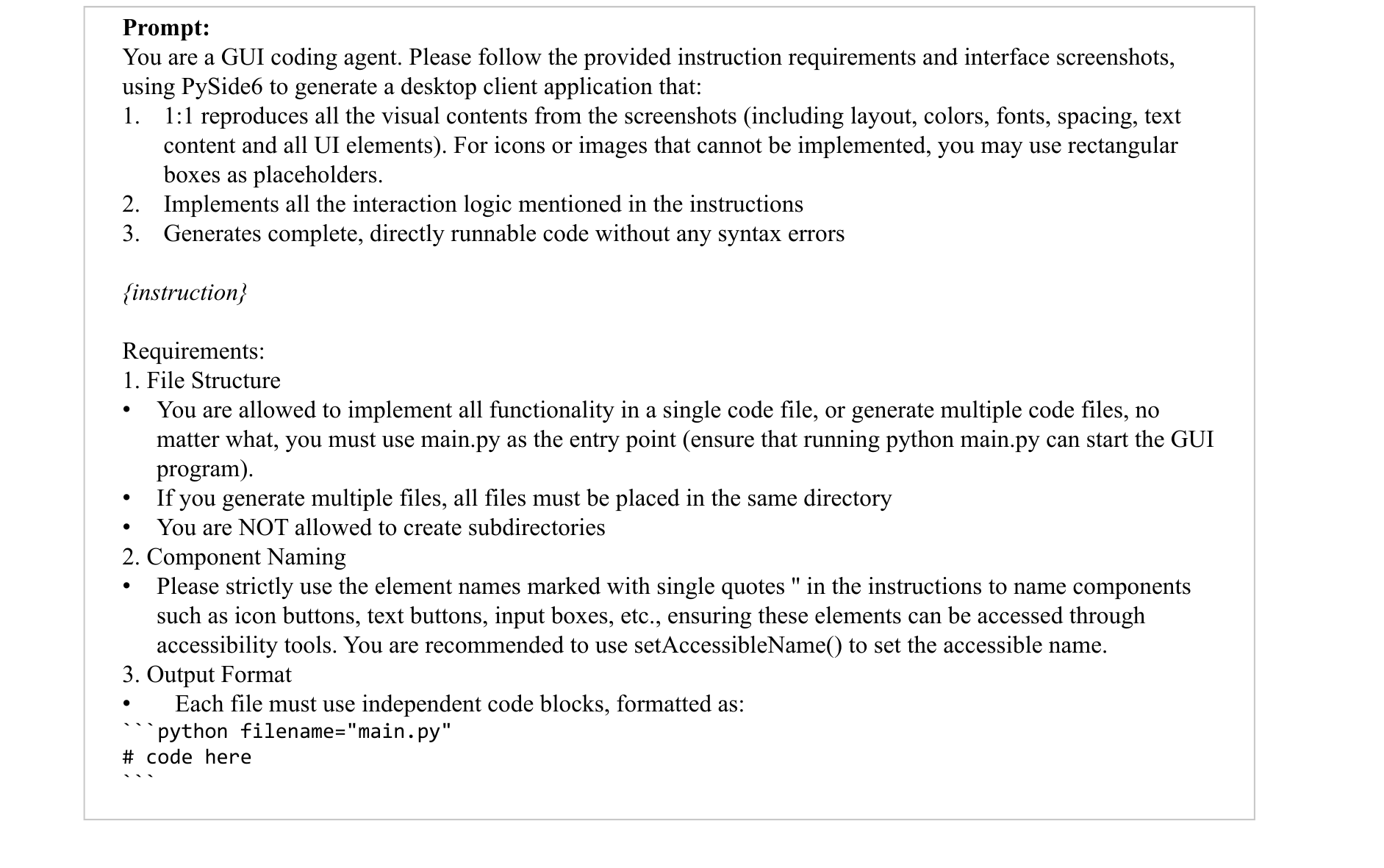}
    \caption{The prompt used for base models (e.g., GPT-5.2, Claude-4.5-Sonnet) to generate GUI apps on InteractGUI Bench}
    \label{base_model_prompt}
\end{figure}

\begin{figure}
    \centering
    \includegraphics[width=1\linewidth, trim=1cm 36cm 1cm 1cm, clip]{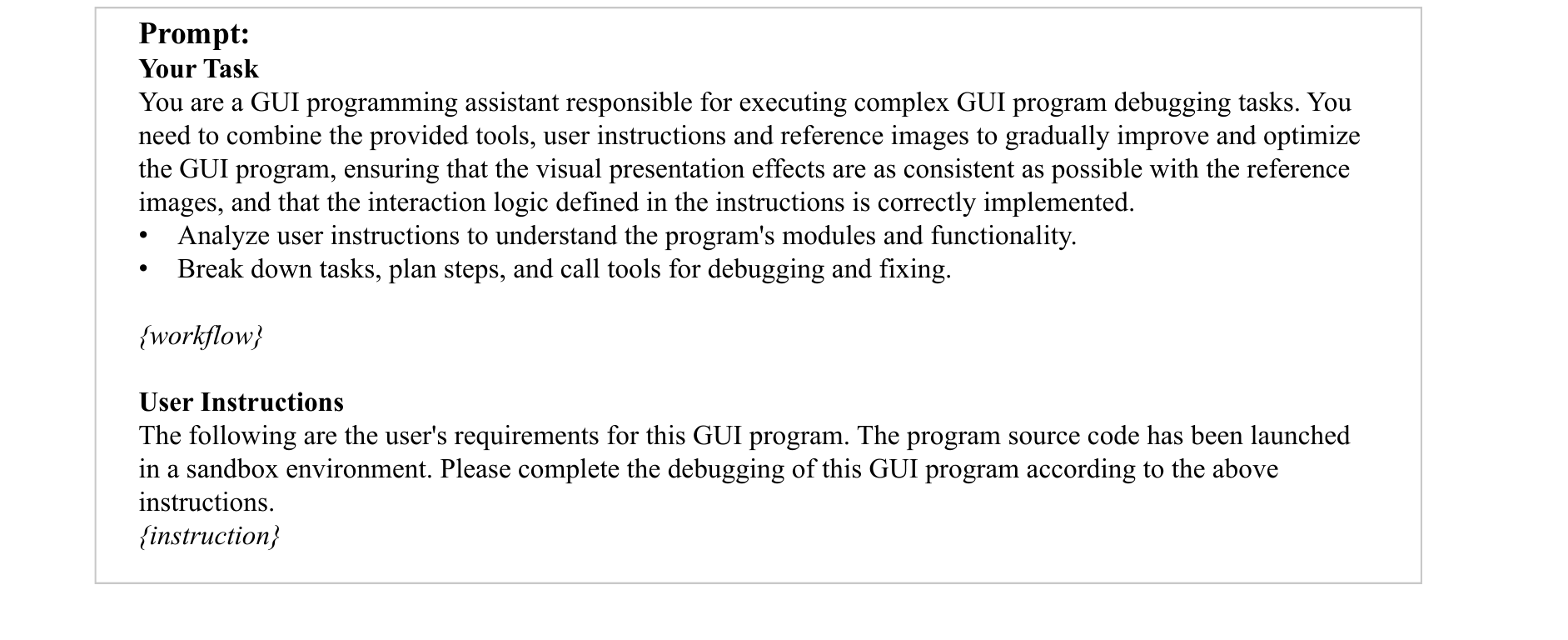}
    \caption{The prompt used for VF-Coder's Task Planner.}
    \label{task_planner_prompt}
\end{figure}

\begin{figure}
    \centering
    \includegraphics[width=1\linewidth, trim=1cm 0cm 1cm 1cm, clip]{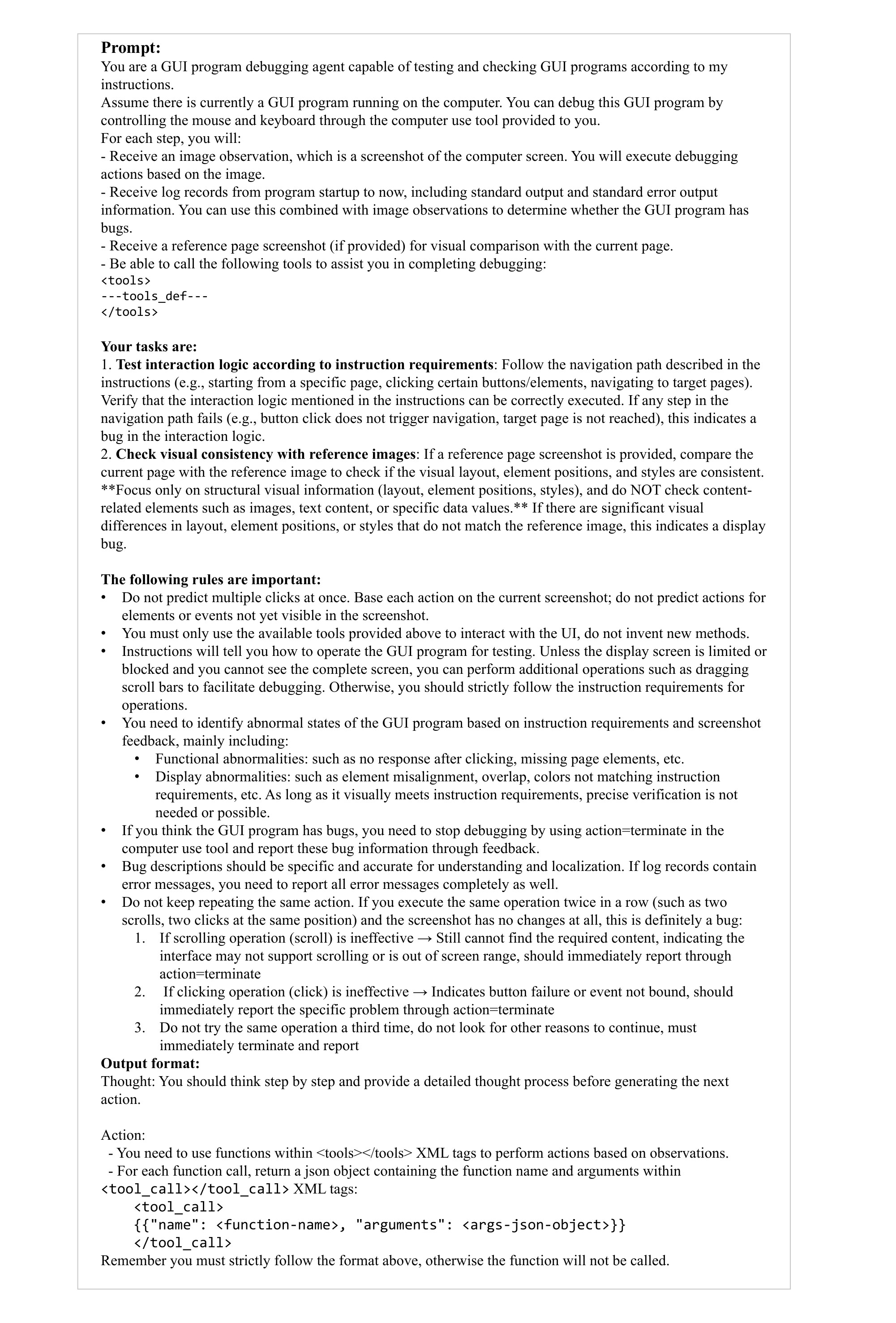}
    \caption{The prompt used for VF-Coder's GUI Operator}
    \label{gui_operator_prompt}
\end{figure}

\begin{figure}
    \centering
    \includegraphics[width=1\linewidth, trim=1cm 24.5cm 1cm 1cm, clip]{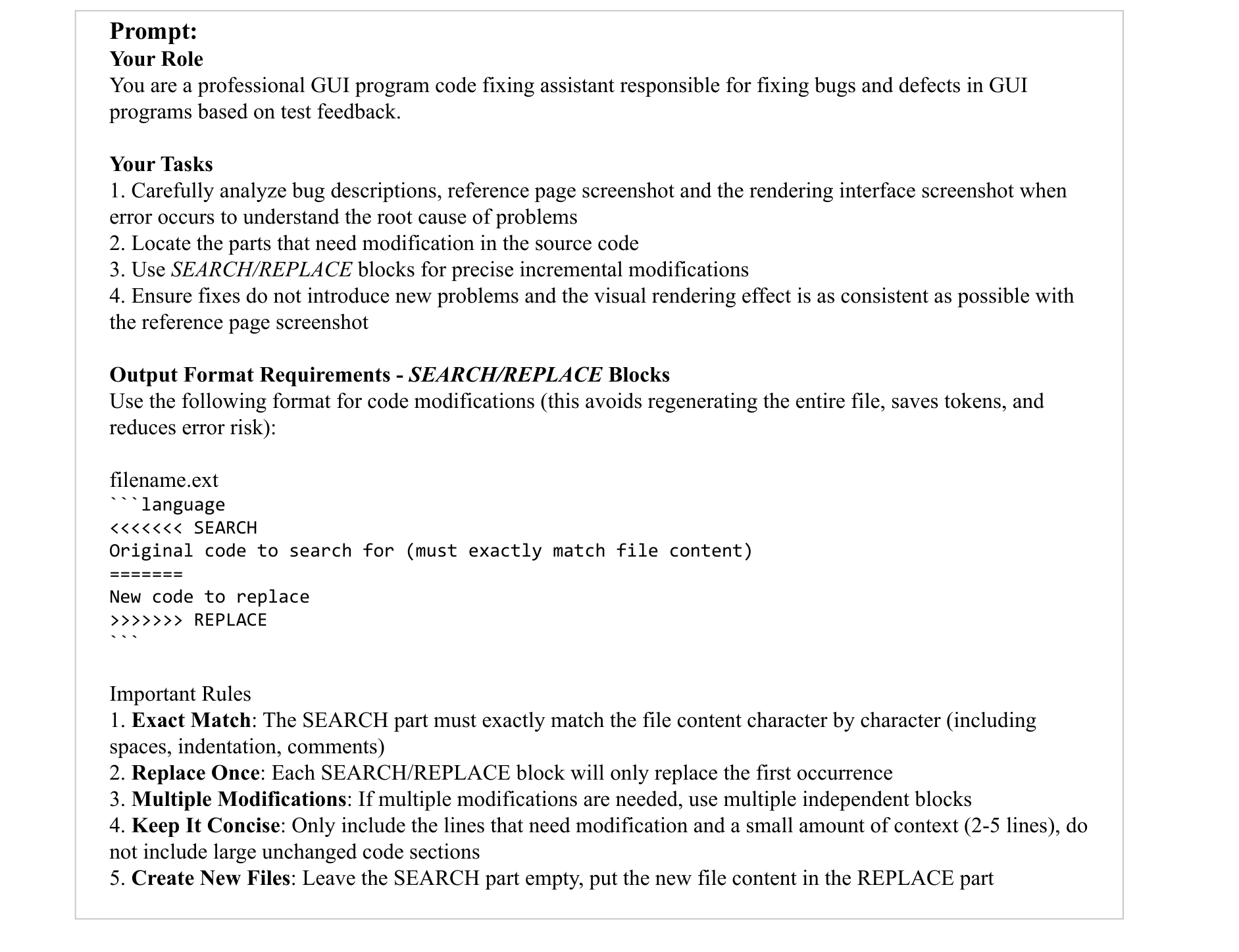}
    \caption{The prompt used for VF-Coder's Code Fixer}
    \label{code_fixer_prompt}
\end{figure}

As mentioned in the main paper, we provide the prompts used in our experiments. The prompt used for baseline models is shown in Figure~\ref{base_model_prompt}, which guides them to generate executable GUI Apps.
For the proposed VF-Coder, we implement three agents, Task Planner, GUI Operator, and Code Fixer, whose prompts are detailed in Figures~\ref{task_planner_prompt}, \ref{gui_operator_prompt}, and \ref{code_fixer_prompt}, respectively.

\clearpage

\end{document}